\documentclass[12pt]{article}
\usepackage[dvips]{color}
\usepackage{amsmath}
\usepackage{graphicx}
\usepackage{subfigure}
\usepackage{epsfig}
\usepackage{amsmath}
\usepackage{cite}
\usepackage{color}
\usepackage{subfigure}
\def\Box{\hbox{$\rlap{$\sqcup$}\sqcap$}}
\def\Box{\hbox{$\rlap{$\sqcup$}\sqcap$}}
\begin{document}
\begin{center}
\Large{\bf Constraints on cosmological parameters in light of the scalar-tensor theory of gravity and swampland conjectures}\\
\small \vspace{1cm}{\bf S. Noori Gashti$^{\star, \dagger}$\footnote {Email:~~~saeed.noorigashti@stu.umz.ac.ir}},
 {\bf J. Sadeghi$^{\star}$\footnote {Email:~~~pouriya@ipm.ir}} \quad
\\
\vspace{0.5cm}$^{\star}${Department of Physics, Faculty of Basic
Sciences,\\
University of Mazandaran
P. O. Box 47416-95447, Babolsar, Iran}\\
$^{\dagger}${School of Physics, Damghan University, P. O. Box 3671641167, Damghan, Iran
\small \vspace{1cm}
}
\end{center}
\begin{abstract}
In this paper, we study various cosmological parameters and quantities in scalar-tensor gravity from inflation
and swampland conjecture. Therefore, by selecting different models as power-law,
exponential, and logarithmic
in the framework of scalar-tensor theory, we obtain potential, tensor-to-scalar ratio, and the scalar
spectral index. Next, we examine new constraints and compare the corresponding results with
the latest observable data. Here,  we take advantage of the obtained results and
determine the compatibility or incompatibility of the corresponding model with the swampland conjectures.
\end{abstract}
\newpage
\tableofcontents
\section{Introduction}

As we know,  researchers consider various inflationary models as a profound concept to explain the structures of the present universe\cite{1,2}. The standard model of cosmology has a series of principles, such as homogeneity in the FLRW framework. It can explain many problems, such as expanding the universe and cosmic microwave background radiation (CMBR). We note here that this model always had problems, such as singularity, flatness,  horizon, structure formation, and dark energy. To solve these problems, researchers have used various solutions in the past decades\cite{3,4,5,6,7,8}.
 For example, by modifying the theory of general relativity, significant efforts have been made to solve these problems. All of these concepts are consistent with recent observable data. The inflation hypothesis shows an accelerated period of expansion in the early universe, and during an inflationary period, a dS universe is driven by cosmological constant\cite{9,10,11,12,13,14,15}. So far, researchers have introduced various types of inflation models. The most famous is the $R^{2}$ Strabiansky model. Different types of modified theories of gravity as $f(R),$ $f(R, T)$, and $f(Q)$ are used to explain the structures of the universe. Inflation is also combined with various conjectures, conditions and their cosmology concept has been studied, which can include the slow-roll, constant-roll and etc\cite{16,17,18,19,20,21,22,23,24,25,26}. On the other hand, non-minimal scalar-tensor theory has been comprehensively studied for cosmological applications from the very early stages of inflation to the last stages via a radiation-dominated era. It has shown that such a particular theory has accepted a suitable inflation regime. We note here the cosmological parameters such as the tensor-to-scalar ratio and the scalar spectral index always have allowable values consistent with observable data. Based on observable data and studies in this context, it seems that the non-minimal coupled to a scalar-tensor
gravitational theory is a very suitable model for describing the history of universe evolution \cite{27,28,29,30,31,32,33,34,35,36,37,38,39,40,41}. Given the above concepts, we will study a specific type of inflationary scenario that has not been studied in the literature. We want to check the inflation of the scalar-tensor gravity model from swampland conjectures. On the other hand, the weak gravity conjecture has two known concepts as swampland and landscape. These conjectures are used to explain the quantum gravity \cite{42,43,44,45,46,47,48,49,50,51,52,53,54,55, 56,57,58,59,60,61,62,63,64,65,66,67}.
By selecting different inflation models, we want to study various cosmological parameters and quantities in scalar-tensor gravity from inflation and swampland conjecture. Therefore, we organized
the continuation of this article in the following form: In section 2, we review the model and introduce some concepts of scalar-tensor gravity. In section 3, we present the cosmological parameters in the concept of scalar-tensor gravity.
In section 4, we calculate cosmological parameters as scalar spectral index, tensor-to-scalar ratio and potential. In that case, we select different inflation models, as power-law,  exponential and logarithmic forms, in the framework of scalar-tensor gravity.
Then we will call the swampland conjectures and combine them with the scalar-tensor gravity in inflation. We will examine new constraints and compare the results of these studies with the
latest observable data, such as Planck 2018, which has imposed new tighter restrictions on cosmological parameters. Here also, we analyze the results and determine the compatibility or incompatibility of this inflation model with the swampland conjectures. And finally, in section 5, we present the result of the corresponding model.

\section{Overview of scalar-tensor gravitational model {STG}}
In this section, we try to introduce and review the scalar-tensor gravitational action
with corresponding equations.
So, the non-minimally coupled scalar-tensor gravity can be expressed by the following
action,
\begin{equation}\label{1}
S=\int d^{4}x\sqrt{-g}\bigg[f(\phi)R-\frac{\omega(\phi)}{\phi}\phi_{,\mu}\phi^{,\mu}-V(\phi)-\mathcal{L}_{m}\bigg],
\end{equation}
where $\mathcal{L}_{m}$, $f(\phi)$ and $\omega(\phi)$ are the matter Lagrangian density,
coupling parameter,
and Brans-Dicke parameter, respectively. So the field equations
are given by,
\begin{equation}\label{2}
\bigg(R_{\mu\nu}-\frac{1}{2}g_{\mu\nu}R\bigg)f(\phi)+g_{\mu\nu}\Box f(\phi)-f_{;\mu;\nu}-\frac{\omega(\phi)}{\phi}\phi_{,\mu}\phi_{,\nu}+\frac{1}{2}g_{\mu\nu}\bigg(\phi_{,\alpha}\phi^{,\alpha}+V(\phi)\bigg)=T_{\mu\nu},
\end{equation}
and
\begin{equation}\label{3}
Rf'+2\frac{\omega(\phi)}{\phi}\Box \phi+\bigg(\frac{\omega'(\phi)}{\phi}-\frac{\omega(\phi)}{\phi^{2}}\bigg)\phi_{,\mu}\phi^{,\mu}-V'(\phi)=0,
\end{equation}

where $\Box$ is D’Alembertian as $\Box f(\phi)=f''\phi_{,\mu}\phi^{,\mu}-f'\Box\phi$.
In general, the coupling parameter, Brans-Dicke parameter, etc.,  is generally selected to investigate the universe's evolution. However, the general conserved current and the above field equations lead us to relate these parameters\cite{37,38,39,40,41}. Hence,  according to equation (2), we will have,
 \begin{equation}\label{4}
Rf-3\Box f-\frac{\omega(\phi)}{\phi}\phi_{,\mu}\phi^{,\mu}-2V=T^{\mu}_{\mu}=T,
\end{equation}
Now with respect to equations (3) and (4), one can obtain following equation,
\begin{equation}\label{5}
\bigg(3f'^{2}+\frac{2\omega f}{\phi}\bigg)'\phi_{,\mu}\phi^{,\mu}+\bigg(3f'^{2}+\frac{2\omega f}{\phi}\bigg)\Box\phi+2f'V-fV=f'T,
\end{equation}
The  above equation can also be expressed by the following form,
\begin{equation}\label{6}
\bigg[\bigg(3f'^{2}+\frac{2\omega f}{\phi}\bigg)^{\frac{1}{2}}\phi^{;\mu}\bigg]_{,\mu}-\frac{f^{3}}{2\bigg(3f'^{2}+\frac{2\omega f}{\phi}\bigg)^{\frac{1}{2}}}\bigg(\frac{V}{f^{2}}\bigg)'=\frac{f'}{2\bigg(3f'^{2}+\frac{2\omega f}{\phi}\bigg)^{\frac{1}{2}}}T,
\end{equation}
and finally we will have,
\begin{equation}\label{7}
\bigg(3f'^{2}+\frac{2\omega f}{\phi}\bigg)^{\frac{1}{2}}\bigg[\bigg(3f'^{2}+
\frac{2\omega f}{\phi}\bigg)^{\frac{1}{2}}\phi^{;\mu}\bigg]_{,\mu}-f^{3}\bigg(\frac{V}{f^{2}}\bigg)'=\frac{f'}{2}T^{\mu}_{\mu},
\end{equation}
According to all the above concepts, the  conserved current $J^{\mu}$ can be expressed by\cite{71,72,73},
\begin{equation}\label{8}
J^{\mu}_{;\mu}=\bigg\{\bigg(3f'^{2}+\frac{2\omega f}{\phi}\bigg)^{\frac{1}{2}}\phi^{;\mu}\bigg\}_{;\mu}=0,
\end{equation}
The importance of the above relations and especially derive this conserved current is completely discussed in\cite{71}. The trace-less matter field as $T^{\mu}_{\mu}=T=0$ lead us to the following equation,
\begin{equation}\label{9}
V(\phi)\propto f(\phi)^{2},
\end{equation}
According to the following FRW metric,
\begin{equation}\label{10}
ds^{2}=-dt^{2}+a^{2}(t)\bigg(\frac{dr^{2}}{1-kr^{2}}+r^{2}(d\theta^{2}+\sin^{2}\theta d\phi^{2})\bigg),
\end{equation}
the conserved current (8) is given by,
\begin{equation}\label{11}
\sqrt{\bigg(3f'^{2}+\frac{2\omega f}{\phi}\bigg)}a^{3}\dot{\phi}=\mathcal{C}_{1},
\end{equation}
where $a(t)$ and $\mathcal{C}_{1}$ are the scale factor and integration constant respectively. The relations between coupling parameter and  potential is specified by equation (9), further using the relations and concepts mentioned in\cite{41,71,74,75}, we will have,
\begin{equation}\label{12}
3f'^{2}+\frac{2\omega f}{\phi}=\omega_{0}^{2},
\end{equation}
where $\omega_{0}$ is a constant parameter. Here also,
one can obtain,
\begin{equation}\label{13}
a^{3}\dot{\phi}=\frac{\mathcal{C}_{1}}{\omega_{0}}=\mathcal{C},
\end{equation}
Where $\mathcal{C}$ is a constant parameter. Given the above concept coupling parameters, the potential and the Brans-Dicke parameter can be assumed constant. Next, we assume $\frac{M_{pl}^{2}}{2}=c=1$ and also consider different $f(\phi)$, which fixes the Brans-Dicke parameters and potential. From these functions point of view, we focus all our attention on the inflation in the early vacuum-dominated universe, but we also challenge swampland conjectures. We add a term to the potential to relax the symmetry. So that one of these parameters will act as a constant, which is related to the cosmological constant. To be more consistent with observable data, we also use some constants, which we will explain in more detail in the next section. The number of parameters involved is increased to three $(\omega_{0}, V_{0}, V_{1})$ to advance this work, which is necessary to fit well with the latest observation data.

\section{Cosmological parameters in STG}.

In this section, we provide more details on the scalar-tensor gravity theory and introduce cosmological parameters. Therefore, to study inflation, we consider a hypothesis as $p=0=\rho$. In this situation, the trace of the matter field disappears, and symmetry is held. Hence, action (1) states as the following form,

 \begin{equation}\label{14}
S=\int d^{4}x\sqrt{-g}\bigg[f(\phi)R-\frac{K(\phi)}{2}\phi_{,\mu}\phi^{,\mu}-V(\phi)\bigg],
\end{equation}
According to the conformal transformation $g_{E_{\mu\nu}}=f(\phi)g_{\mu\nu}$,
 Einstein's form of the corresponding action is \cite{68}.
\begin{equation}\label{15}
S=\int d^{4}x\sqrt{-g_{E}}\bigg[R_{E}-\frac{1}{2}(\partial_{E}\sigma)^{2}-V_{E}(\sigma(\phi))\bigg],
\end{equation}

Where $E$ denotes Einstein’s frame. In this framework,
the effective potential $V_{E}$ as well as the
field $\sigma$ can obtain in the following form,
\begin{equation}\label{16}
V_{E}=\frac{V(\phi)}{f^{2}(\phi)},
\end{equation}
and
\begin{equation}\label{17}
(\frac{d\sigma}{d\phi})^{2}=\frac{K(\phi)}{f(\phi)}+3\frac{f'^{2}(\phi)}{f^{2}(\phi)}=\frac{2\omega(\phi)}{\phi f(\phi)}+3\frac{f'^{2}(\phi)}{f^{2}(\phi)},
\end{equation}

In Einstein's frame, according to the action (15), the field equations
can be obtained by,
\begin{equation}\label{18}
\ddot{\sigma}+3H\dot{\sigma}+V'_{E}=0,
\end{equation}
\begin{equation}\label{19}
3H^{2}=\frac{1}{2}\dot{\sigma}^{2}+V_{E},
\end{equation}
where  the above equations, $H$ is expressed by $H=\frac{\dot{a}_{E}}{a_{E}}$.
For cosmological studies, we use some  important parameters as
 slow-roll parameters  and  number of e-folds, which are given by,
\begin{equation}\label{20}
\epsilon=(\frac{V'_{E}}{V_{E}})^{2}(\frac{d\sigma}{d\phi})^{-2},
\end{equation}
\begin{equation}\label{21}
\eta=2\bigg[(\frac{V''_{E}}{V_{E}})(\frac{d\sigma}{d\phi})^{-2}-(\frac{V'_{E}}{V_{E}})(\frac{d\sigma}{d\phi})^{-3}(\frac{d^{2}\sigma}{d\phi^{2}})\bigg],
\end{equation}
 and
\begin{equation}\label{22}
N=\int_{t_{i}}^{t_{f}}Hdt=\frac{1}{2\sqrt{2}}\int_{\phi_{i}}^{\phi_{e}}\frac{d\phi}{\sqrt{\epsilon}}\frac{d\sigma}{d\phi},
\end{equation}

To further explain the above equations, we can add some points, for example,
$t_{i}$ and $t_{f}$ show initiation and the end time, $\phi_{i}$ and $\phi_{e}$
represent the scalar field
at the beginning and end of inflation, respectively. Also, we obtain a
series of cosmological parameters
such as the tensor-to-scalar ratio $r$ and the scalar spectral index $n_{s}$.
In that case we compare the primordial curvature perturbation
$P_{\zeta}(K)$ with the primordial gravitational wave power spectrum $P_{t}(K)$, where
 $r=\frac{P_{t}(K)}{P_{\zeta}(K)}$ and
 $n_{s}-1=2\eta-6\epsilon$. Also, according to the latest observable data
 from TT,TE,EE+lowEB+lensing and TT,TE,EE+lowE+lensing+BK14+BAO,
  different values are mentioned for these parameters. They determined the tensor-to-scalar
  ratio is $r\leq0.16$,  $r\leq0.07$ and the
  scalar spectral index is  $n_{s}\leq 0.9569,  n_{s}\leq 0.9815$  respectively \cite{7}.

\section{Swampland conjectures and scalar-tensor gravity}
We note here if the potential in the Einstein frame is constant, the flat part of the effective potential will still be available\cite{69,70}. As mentioned in the second section, symmetry causes the potential in the Einstein frame to be constant, so we must relax the symmetry as stated in condition (9). So, by adding a fixed parameter as $V_{0}$,  the effective potential in the Einstein frame is asymptotically constant. In this paper, we want to check inflation from the scalar-tensor gravity (symmetry of the non-minimal coupled to the scalar-tensor theory of gravity to fixing quantities and parameters such as Brans-Dicke parameter and the potential) and swampland conjectures point of view. Therefore, we calculate cosmological parameters as scalar spectral index, tensor-to-scalar ratio and potential by selecting different inflation models, i.e., power-law,  exponential and logarithmic. Then we will call the swampland conjectures and combine them with the scalar-tensor gravity in inflation. We will examine new constraints and compare the results of these studies with the latest observable data such as Planck 2018\cite{7}, which has imposed new tighter constraints on cosmological parameters. Finally, we analyze the results to determine the compatibility or incompatibility of this inflation model with the swampland conjectures.\\ Swampland conjectures are used in various types of cosmological studies. In recent decades, researchers have done a great deal of research on these conjectures, also; in this paper, we  use the refined swampland conjecture, namely dS swampland\cite{45,46,47,48,49,50,51,52,53,54,55}, which is given by,
\begin{equation}\label{23}
M_{pl}\frac{|V'|}{V}>C_{1},\hspace{1cm}M_{pl}^{2}\frac{|V''|}{V}<-C_{2},
\end{equation}
Where $ C_ {1} $ and $ C_ {2} $ are constant parameters. We selected several models, i.e.,  the power-law form, logarithmic and exponential form, and combined them with various conditions.
Here, first, we consider a power-law potential
\section{The Models}
To challenge the swampland conjecture from the tensor-scalar gravitational theory perspective, we consider three general models to study the inflation,i.e., we challenge the compatibility of the three issue viz the swampland conjecture, the tensor-scalar gravitational theory and the some important inflationary model by calculating cosmological parameters.  Usually, gravitational corrections are often overlooked in studies related to eternal inflation, and exclusively the large field inflation models are assessed.
 To consider quantum gravitational corrections, we must assume small field models, which are also possible through swampland conjectures.
Of course, there are a few things to keep in mind about swampland conjectures, and the most important is that this idea has not yet been accepted as a complete theory and has not been widely accepted in the literature. However, it has been studied in many cosmological concepts. These conjectures are somewhat consistent with approaches such as warm inflation\cite{aa}.
 Of course, each of the swampland conjectures is constantly being reviewed and completed, and even corrections are made in each of these conjecture to be more compatible with the models and the latest observable data\cite{7}.
However, these conjectures still face serious challenges.
But perhaps in the not-too-distant future, through observations, the compatibility of this idea with theories can be increased,
and it can even find a place among physicists as an accepted idea.
Of course, it is still being studied as a nascent idea in studying many cosmic structures such as inflation and the physics of black holes and dark energy.
The next point is that gravitational corrections are negligible in small field models and potentially predominant in large field models\cite{61}.
For this reason, effective field theories about large fields are not entirely reliable.
Of course, critical points can also be mentioned concerning eternal inflation.
Matsui, Takahashi, and Dimopoulos discussed the swampland conjectures with respect to eternal inflation\cite{76}. They showed the originally proposed de Sitter swampland conjectures to be generically incompatible with eternal inflation. Also, there are several studies about this view under certain conditions. For example, more recently, William H. Kinney showed in his recent article\cite{62}  that the swampland conjecture, which somehow applies weaker criteria to the potential of the scalar field (Refined Swampland Conjecture) in inflation, is somewhat compatible with eternal inflation. Therefore, if the refined conjecture is valid, the presence of a landscape-based multiverse in string theory is not inconsistent with a self-consistent UV completion, and it has remarkable outcomes for model construction in string theory. (For further study, see \cite{76,77,78,79}).
This paper will discuss the compatibility of cosmological parameters in scalar-tensor gravitational theory with swampland conjectures.
Recently, however, Yuennan and Channuie, in an article entitled Further refining Swampland Conjecture on inflation
in general scalar-tensor theories of gravity examined the consistency of extended swampland dS conjecture about the general scalar-tensor theories, the results of which in turn can be interesting\cite{80}.
\subsection{Model I; Power- Law Potential}
In this section, we consider a power-law potential, which is selected by form of $f(\phi)=\phi^{n}$ and the  potential is  expressed by $V(\phi)=\phi^{2n}$. Then we select $n=\frac{1}{2}$ and examine the corresponding model. Here, the main purpose is to study the compatibility or incompatibility of combining the studied model in the scalar-tensor gravity framework with the
swampland conjecture with respect to observable data. In the following, we can briefly review for other powers of $n$ and discuss the overall effect of $n.$ However, by using the above explanations and considering a constant $V_{0}$  for the potential, we will have.

\begin{equation}\label{24}
V(\phi)=V_{1}\phi^{2n}+V_{0},
\end{equation}

According to the equations (12), (16), (17),  (20), (21) and (22) and $n=\frac{1}{2}$,
we calculate the Brans-Dicke parameter,  potential,
the expression for $\frac{d\sigma}{d\phi}$ and slow-roll parameters,
which are given by, \begin{equation}\label{25}
\omega(\phi)=\frac{-3+4\phi \omega_{0}^{2}}{8\sqrt{\phi}},
\end{equation}
 \begin{equation}\label{26}
V_{E}=V_{1}+V_{0}\phi^{-1},
\end{equation}

\begin{equation}\label{27}
\frac{d^{2}\sigma}{d\phi^{2}}=\frac{\omega_{0}^{2}}{\phi},
\end{equation}

\begin{equation}\label{28}
\epsilon=\frac{V_{0}^{2}\phi}{\omega_{0}^{2}(V_{0}\phi+V_{1}\phi^{2})^{2}},
\end{equation}
and
\begin{equation}\label{29}
\eta=\frac{3V_{0}}{\phi\omega_{0}^{2}(V_{0}+V_{1}\phi)},
\end{equation}
In the above equations, if we assume the value of the constant potential $V_{0}$ set to be zero, the effective potential in the Einstein frame will be equal to $V_{E}=V_{1}$ and it remains flat. Now we will use the swampland conjectures in the corresponding model and examine its compatibility with observable data. In that case, according to equation (26), we have  to obtain the first and second derivatives of the effective potential in terms of $\phi$ so that we will have,

\begin{equation}\label{30}
V'(\phi)=-\frac{V_{0}}{\phi^{2}}, \hspace{12pt}V''(\phi)=\frac{2V_{0}}{\phi^{3}},
\end{equation}

Now with respect to equations (23), (26), and (30), one can calculate,

\begin{equation}\label{31}
\frac{-\frac{V_{0}}{\phi^{2}}}{V_{1}+V_{0}\phi^{-1}}>C_{1}, \hspace{12pt}\frac{\frac{2V_{0}}{\phi^{3}}}{V_{1}+V_{0}\phi^{-1}}<-C_{2},
\end{equation}
As we know, the tensor-to-scalar ratio $r$ and scalar spectral index $n_{s}$ are important quantities in cosmology.  So, according to slow-roll parameters in equations (28) and (29),
we can have the following equations,

\begin{equation}\label{32}
n_{s}=1-\frac{6V_{0}^{2}\phi}{\omega_{0}^{2}(V_{0}\phi+V_{1}\phi^{2})^{2}}+\frac{6V_{0}}
{\phi\omega_{0}^{2}(V_{0}+V_{1}\phi)},
\end{equation}
 and
\begin{equation}\label{33}
r=\frac{16V_{0}^{2}\phi}{\omega_{0}^{2}(V_{0}\phi+V_{1}\phi^{2})^{2}},
\end{equation}

As mentioned in the previous section, observable data has created new constraints on cosmological parameters such as $n_{s}$ and $r$.  So, by selecting the different values for free parameters as $(V_{1}, V_{0}, \omega_{0})$, the important cosmological parameters obtain different values compared with the latest observable data. According to the component $f(\phi)$ also $\phi_{i}=1.9$, $V_{0}=0.9\times 10^{-13}T^{-2}$ and $V_{1}=0.9\times 10^-{13}T^{-2}$ it can be checked that different values obtained for the parameters under consideration. Hence we will have $10<\omega_{0}<30$, $1.0124<\phi_{e}<1.0315$, $0.000215<r<0.00945$, $0.9705<n_{s}<0.9922$ and $45<N<64$.  Also this rout can be studied for different values of the parameter $n$ according to the $f(\phi)$. As shown in equations (32) and (33), the two cosmological parameters are expressed in the scalar field $\phi$. Now, suppose we reverse these two equations so that the scalar field $\phi$ is expressed in terms of two parameters $n_{s}$ and $r$ and then placed in equation (31). In that case, we can obtain the coefficients of the swampland conjectures in terms of these two cosmological parameters. By plotting some figures and according to the coefficients of the swampland conjectures, we show the changes of these two
cosmological parameters, also we can compare them with the latest observable data. Therefore, to show these variations, we will plot the figures $C_{1, 2}-n_{s}$ and $C_{1, 2}-r$, respectively.

\begin{figure}[h!]
\begin{center}
\subfigure[]{
\includegraphics[height=6cm,width=6cm]{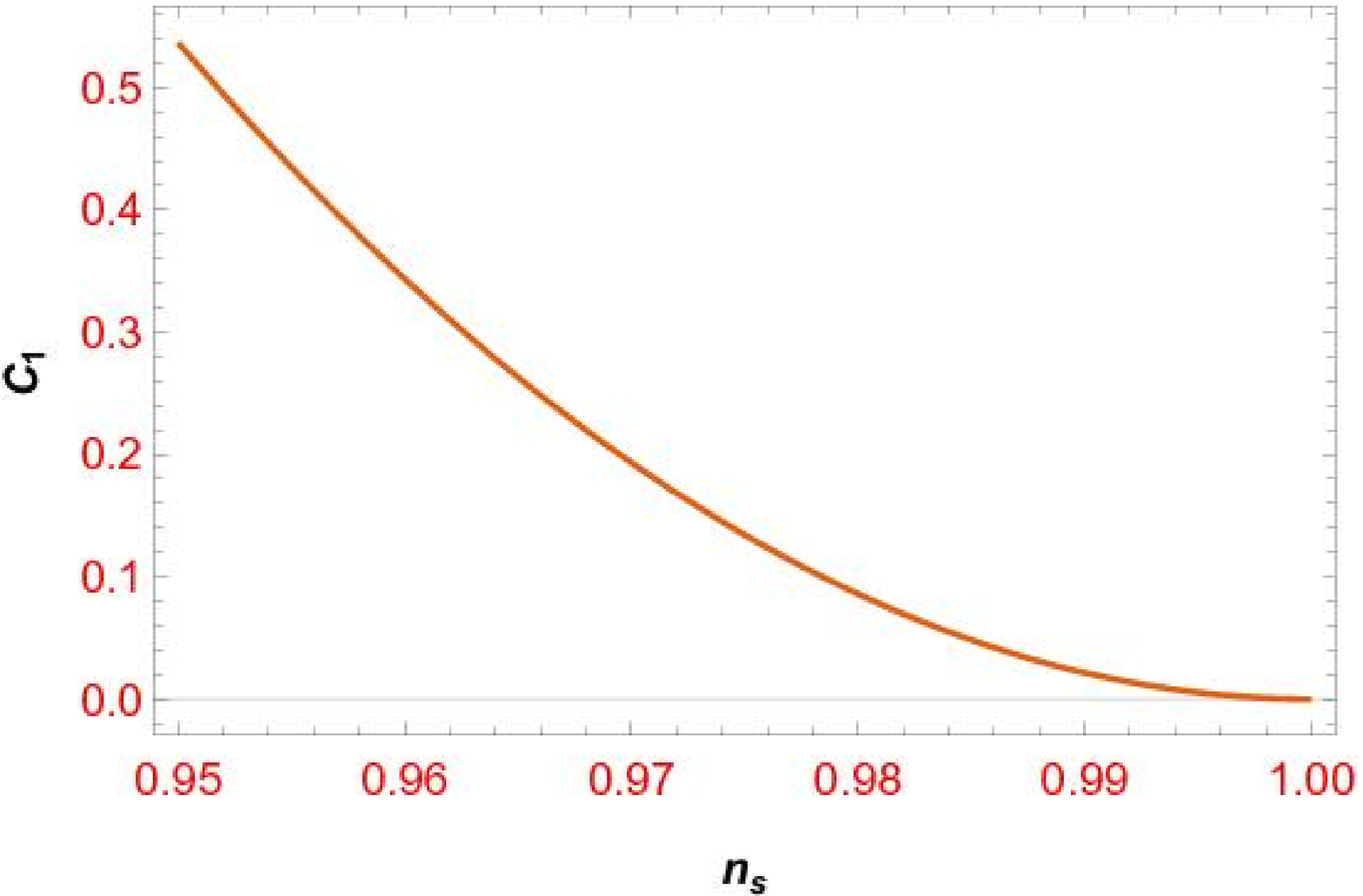}
\label{1a}}
\subfigure[]{
\includegraphics[height=6cm,width=6cm]{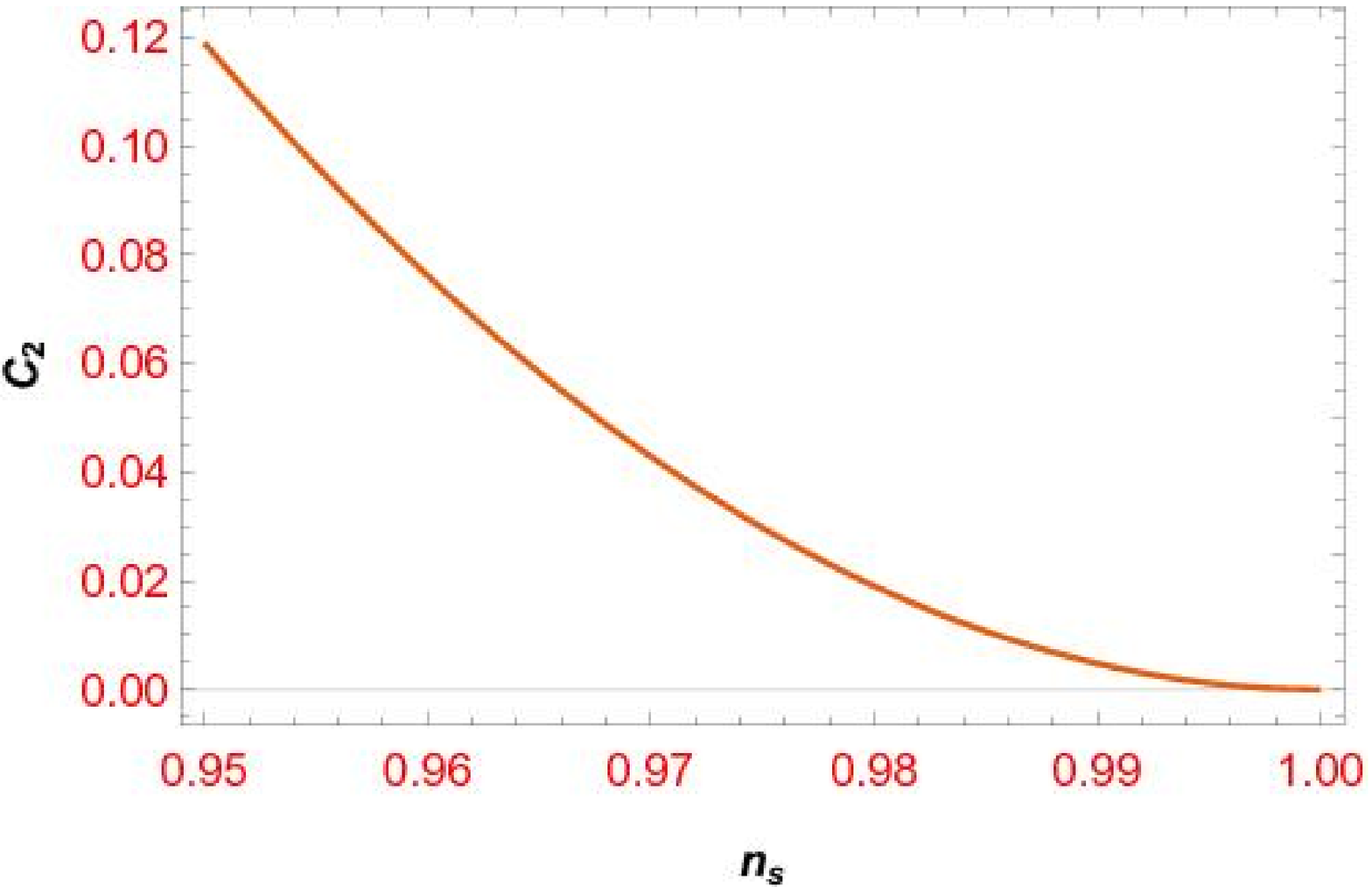}
\label{1b}}
\caption{\small{The plot of $C_{1, 2}$ in terms of $n_{s}$ with respect to free parameter values as $(V_{1}=-0.9\times10^{-13},  V_{0}=0.9\times10^{-13}, \omega_{0}=13 )$ }}
\label{1}
\end{center}
\end{figure}

\begin{figure}[h!]
\begin{center}
\subfigure[]{
\includegraphics[height=6cm,width=6cm]{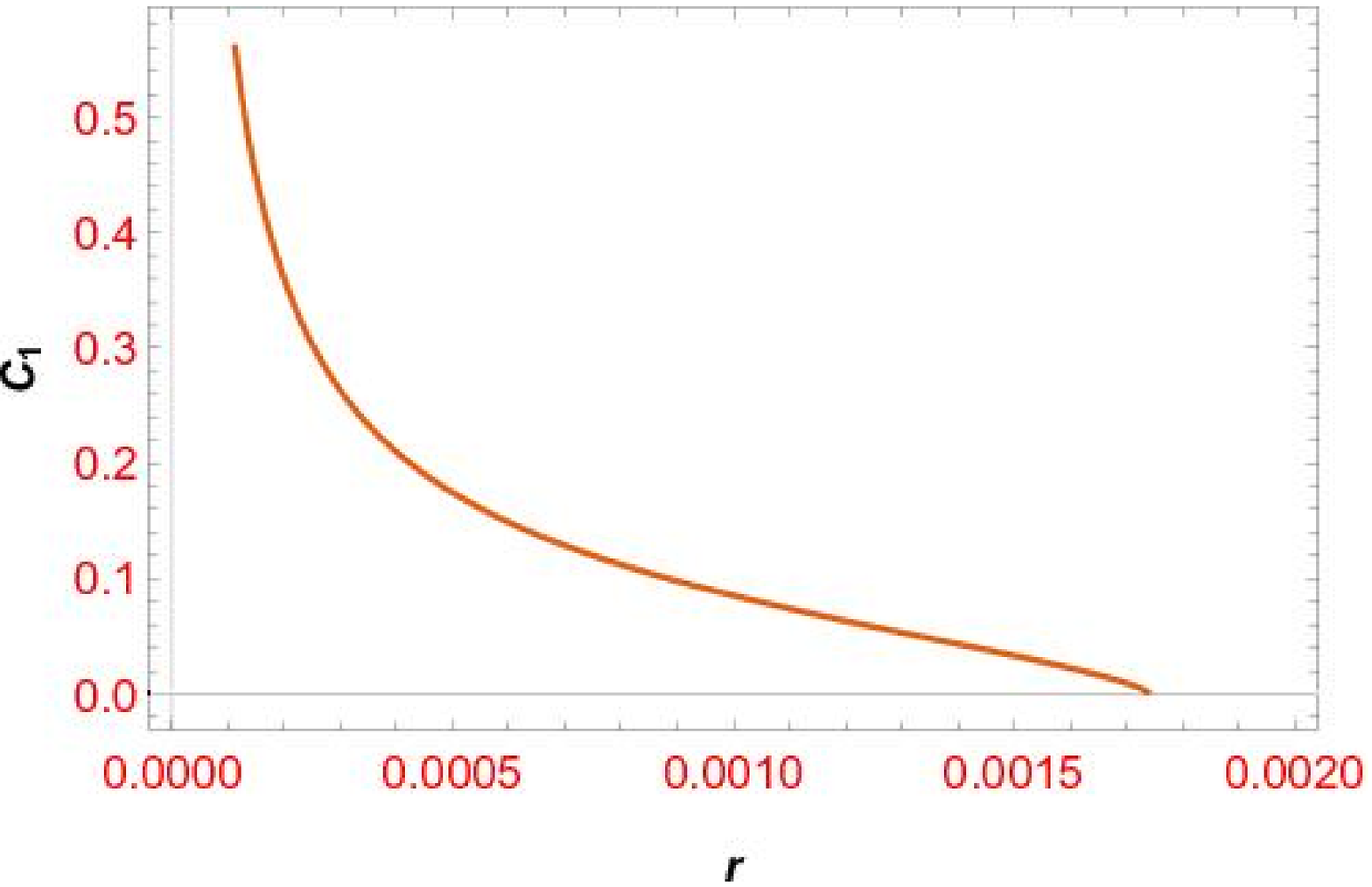}
\label{2a}}
\subfigure[]{
\includegraphics[height=6cm,width=6cm]{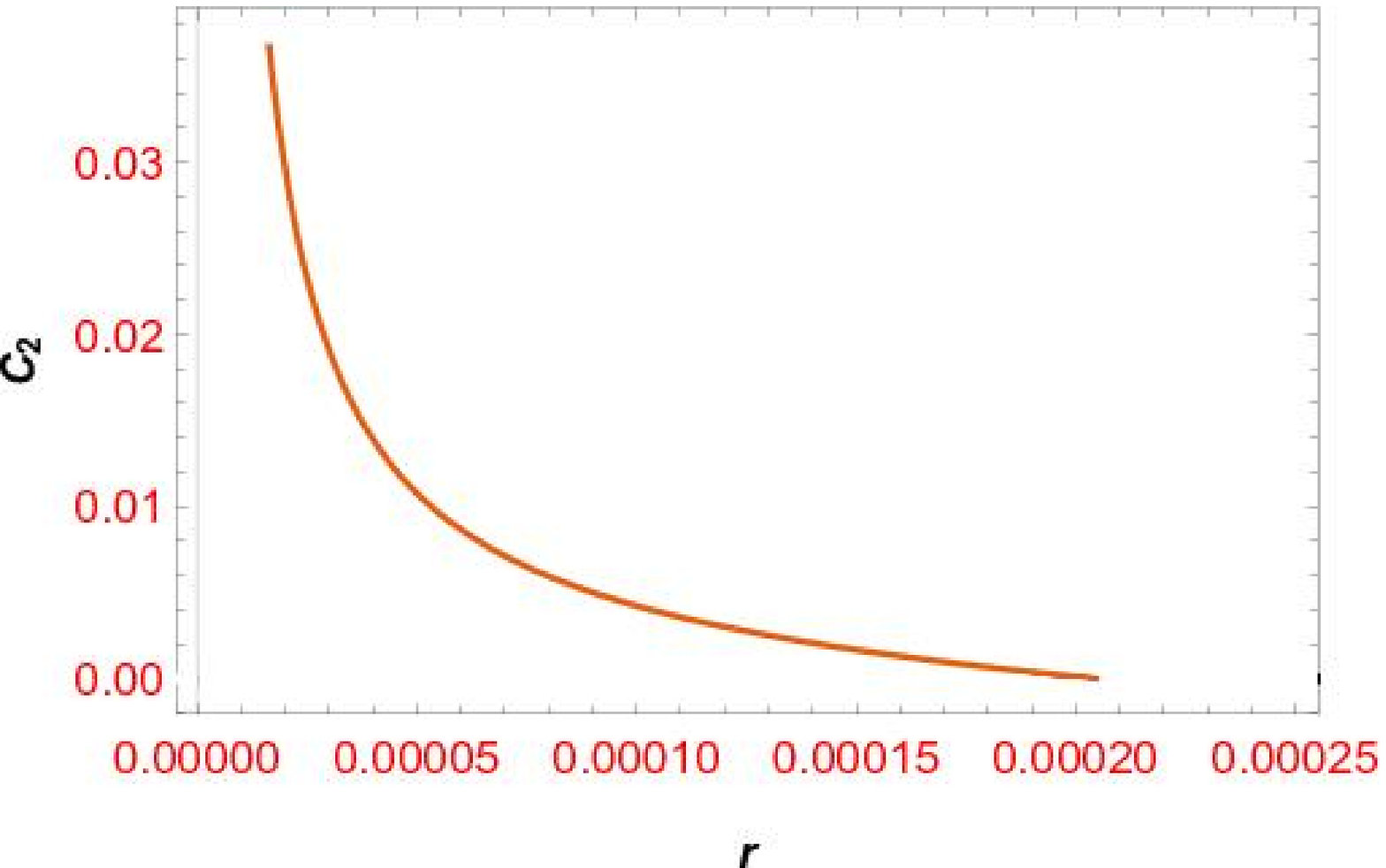}
\label{2b}}
\caption{\small{The plot of $C_{1, 2}$ in terms of $r$ with respect to free parameter values as $(V_{1}=-0.9\times10^{-13},  V_{0}=0.9\times10^{-13}, \omega_{0}=13 )$ }}
\label{2}
\end{center}
\end{figure}

\begin{figure}[h!]
 \begin{center}
 \subfigure[]{
 \includegraphics[height=6cm,width=6cm]{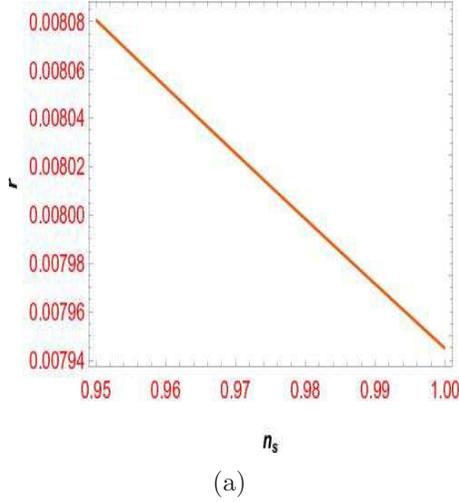}
 \label{3}}
 \caption{The plot of $r$ in term of $n_{s}$ with respect to the free parameter values as $(V_{1}=-0.9\times10^{-13},  V_{0}=0.9\times10^{-13}, \omega_{0}=13 )$}
 \label{3}
 \end{center}
 \end{figure}
As mentioned before, the swampland conjectures have been rewritten in terms of two cosmological parameters $n_{s}$ and $r$.  Here,  by using different figures, we explain the constraints on cosmological parameters, which are described by following: As shown in figure (1),  we plotted the constraints of the first and second components of the swampland conjecture in terms of the scalar spectral index concerning the free parameters as $(V_{0}, V_{1}, \omega_{0})$. The associated range of the scalar spectral index which is found here must be consistent with Planck's observable data\cite{7}. Also, the range of swampland conjecture components is well defined and usually specified by the literature in order 0.1 and 1. Also, as shown in figure (1), the coefficient of $C_{2}$ has smaller than the $C_{1}$. Similarly, in figure (2), new constraints are examined,  namely the constraint of swampland components to tensor-to-scalar ratio. As shown in figure (2), the range associated with $r$ and the swampland conjectures coefficient are also well defined, in which case the values are consistent with the observable data\cite{7}. To plot these figures, we also used free parameters whose values we had determined. In figure(3), we have plotted the restrictions of two cosmological parameters with respect to each other, and the allowable range of these two parameters is well shown in figure (3). As observed in the calculations and diagrams of this section, the model of power-Law potential in combination with the scalar-tensor theory of gravity and swampland conjectures is consistent with the observable data. Also, here we see that the scalar-tensor theory of gravity can also be well compatible with swampland conjectures. We determine the allowable range $C_{i}$ of this model according to the explanations mentioned earlier and the swampland conjecture in terms of cosmological parameters. Therefore, with respect to the components of swampland conjecture, i.e., $C_{1}$ and $C_{2}$ as well as the constant parameters mentioned in this paper and the observable values for $r$ and $n_{s}$, a relation can be determined as $f\geq C_{1}^{2}C_{2}^{2}$. As a result, the allowable range of swampland conjectures for such a model is well illustrated in figure (4). As mentioned concept specified the allowable constraints of $C_{i}$  in figure(4). In the continuation of the article, we will determine the allowable $C_{i}$ for other models according to the cosmological parameters mentioned in the article. In that case, the best model can determine by the above concepts.

\begin{figure}[h!]
 \begin{center}
 \subfigure[]{
 \includegraphics[height=6cm,width=6cm]{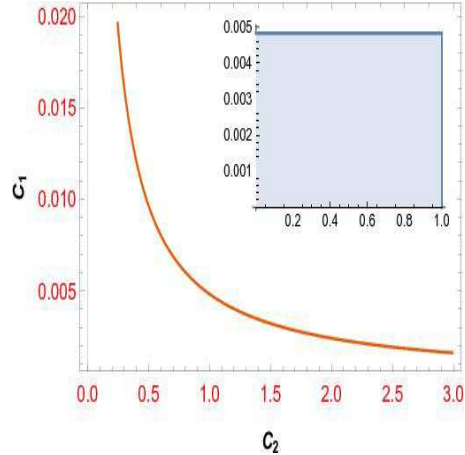}
 \label{4}}
 \caption{The constraints for $C_{1}$, and $C{2}$ with respect to the free parameter values as $(V_{1}=-0.9\times10^{-13},  V_{0}=0.9\times10^{-13}, \omega_{0}=13 )$ and tensor-to-scalar ratio $r$ and scalar spectral index $n_{s}$}
 \label{3}
 \end{center}
 \end{figure}

\subsection{Model II; exponential Potential}

Like the pervious section, here  we consider a exponential potential.
 By selecting  $f(\phi)=\exp(\frac{n\phi}{2})$,
 the potential will be as  $V(\phi)=\exp(n\phi)$. Then we select $n=1$ and consider
  also constant of $V_{0}$ for the potential, we will have,
\begin{equation}\label{34}
V(\phi)=V_{1}\exp(\phi)+V_{0},
\end{equation}

By using equations (12), (16), (17),  (20), (21), and (22), we will calculate the Brans-Dicke parameter and corresponding  potential, which are given by,

\begin{equation}\label{35}
\omega(\phi)=\frac{(\omega_{0}^{2}-\frac{3}{4}\exp(\phi))\times\phi}{2\exp(\frac{\phi}{2})},
\end{equation}
 \begin{equation}\label{36}
V_{E}=V_{1}+V_{0}\exp(-\phi),
\end{equation}

As we know,  the expression for $\frac{d\sigma}{d\phi}$ will be as,
\begin{equation}\label{37}
\frac{d^{2}\sigma}{d\phi^{2}}=\frac{\omega_{0}^{2}}{\exp(\phi)},
\end{equation}

and also, the  slow-roll parameters in Einstein’s frame  are calculated by the following form,

\begin{equation}\label{38}
\epsilon=\frac{V_{0}^{2}\exp(\phi)}{\omega_{0}^{2}(V_{0}+V_{1}\exp(\phi))^{2}},
\end{equation}
 and
\begin{equation}\label{39}
\eta=\frac{V_{0}\exp(\phi)}{\omega_{0}^{2}(V_{0}+V_{1}\exp(\phi))},
\end{equation}

Now we want to use the swampland conjectures and to examine its compatibility with observable data.  So, we need to obtain the first and second derivatives of the effective potential in terms of $\phi$. According to equation (36), we will have,

\begin{equation}\label{40}
V'(\phi)=-V_{0}\exp(-\phi), \hspace{12pt}V''(\phi)=V_{0}\exp(-\phi),
\end{equation}

Now we use the equations (23), (36), and (40), we calculate the following expressions,

\begin{equation}\label{41}
\frac{-V_{0}\exp(-\phi)}{V_{1}+V_{0}\exp(-\phi)}>C_{1}, \hspace{12pt}\frac{V_{0}\exp(-\phi)}{V_{1}+V_{0}\exp(-\phi)}<-C_{2},
\end{equation}
Also, we note here, by considering the equations (38) and (39) the
tensor-to-scalar ratio $r$ and scalar spectral index $n_{s}$,  can  be obtained by the following expressions,

\begin{equation}\label{42}
n_{s}=1-\frac{6V_{0}^{2}\exp(\phi)}{\omega_{0}^{2}(V_{0}+V_{1}\exp(\phi))^{2}}+\frac{2V_{0}\exp(\phi)}{\omega_{0}^{2}(V_{0}+V_{1}\exp(\phi))},
\end{equation}
 and
\begin{equation}\label{43}
r=\frac{16V_{0}^{2}\exp(\phi)}{\omega_{0}^{2}(V_{0}+V_{1}\exp(\phi))^{2}},
\end{equation}
As mentioned before, the observable data has created new constraints on cosmological parameters such as $n_{s}$ and $r$.  So we consider the free parameters as $V_{1}$, $V_{0}$, $\omega_{0}$. According to the component $f(\phi)$ also $\phi_{i}=1.2$, $V_{0}=1.1\times 10^{-20}T^{-2}$ and $V_{1}=1.1\times 10^{-20}T^{-2}$ it can be checked that different values obtained for the parameters under consideration. Hence we will have $10<\omega_{0}<30$, $0.02589<\phi_{e}<0.08569$, $0.015<r<0.065$, $0.9615<n_{s}<0.9842$ and $30<N<78$.  Also this rout can be studied for differen $f(\phi)$. We see $n_{s}$ and $r$ are expressed in terms of the scalar field $\phi$ in equations (42) and (43). Now, like the previous part, if we reverse these two equations so that the scalar field $\phi$ is expressed in terms of two parameters $n_{s}$ and $r$ and then placed in equation (41),
we can obtain the coefficients of the swampland conjectures in terms of these two cosmological parameters. By plotting some figures and comparing the results with observable data, it is possible to show the changes of these two cosmological parameters,  according to the coefficients of the swampland conjectures. So, to show these variations, we will plot the figures $C_{1, 2}-n_{s}$ and $C_{1, 2}-r$, respectively.
\begin{figure}[h!]
\begin{center}
\subfigure[]{
\includegraphics[height=6cm,width=6cm]{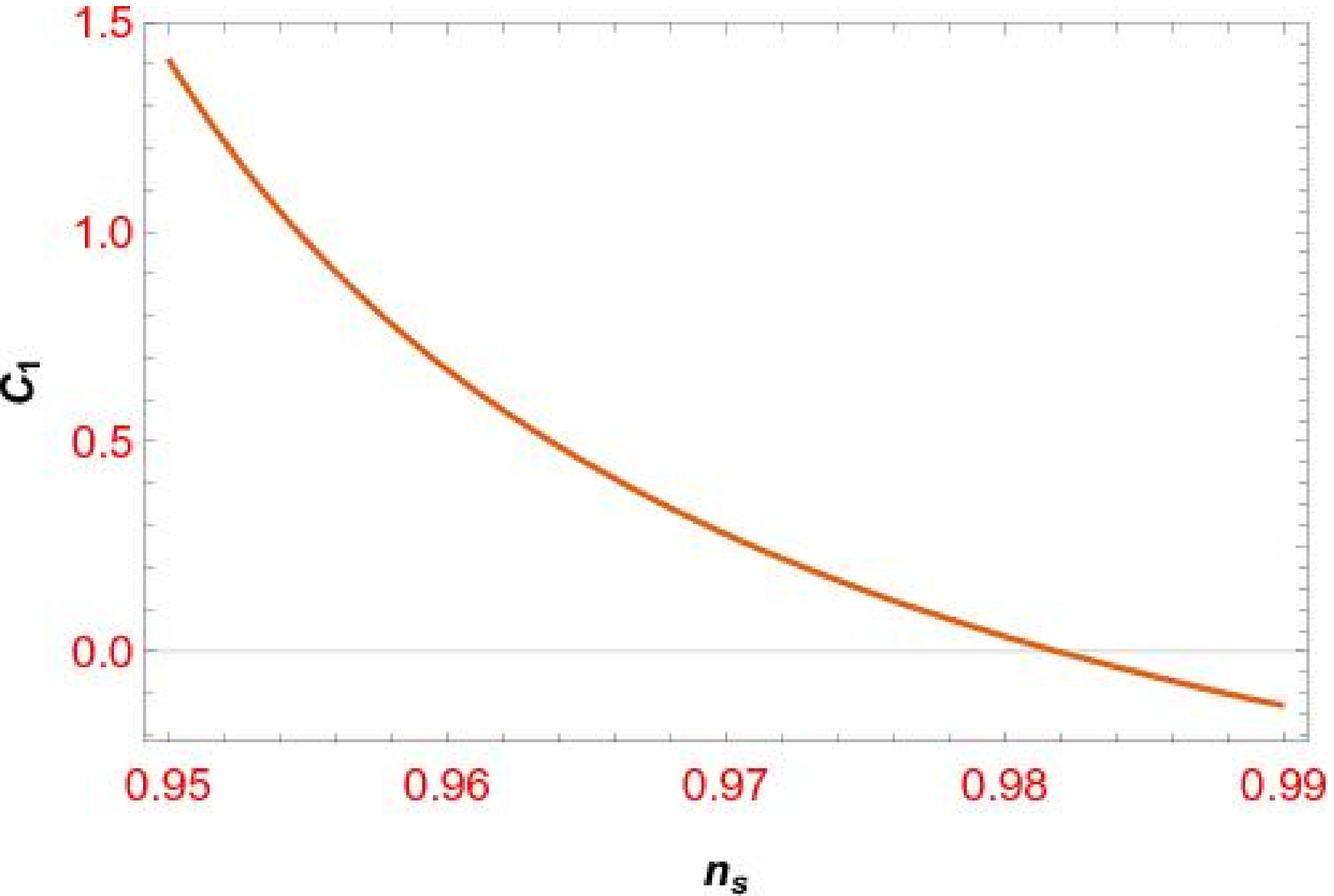}
\label{5a}}
\subfigure[]{
\includegraphics[height=6cm,width=6cm]{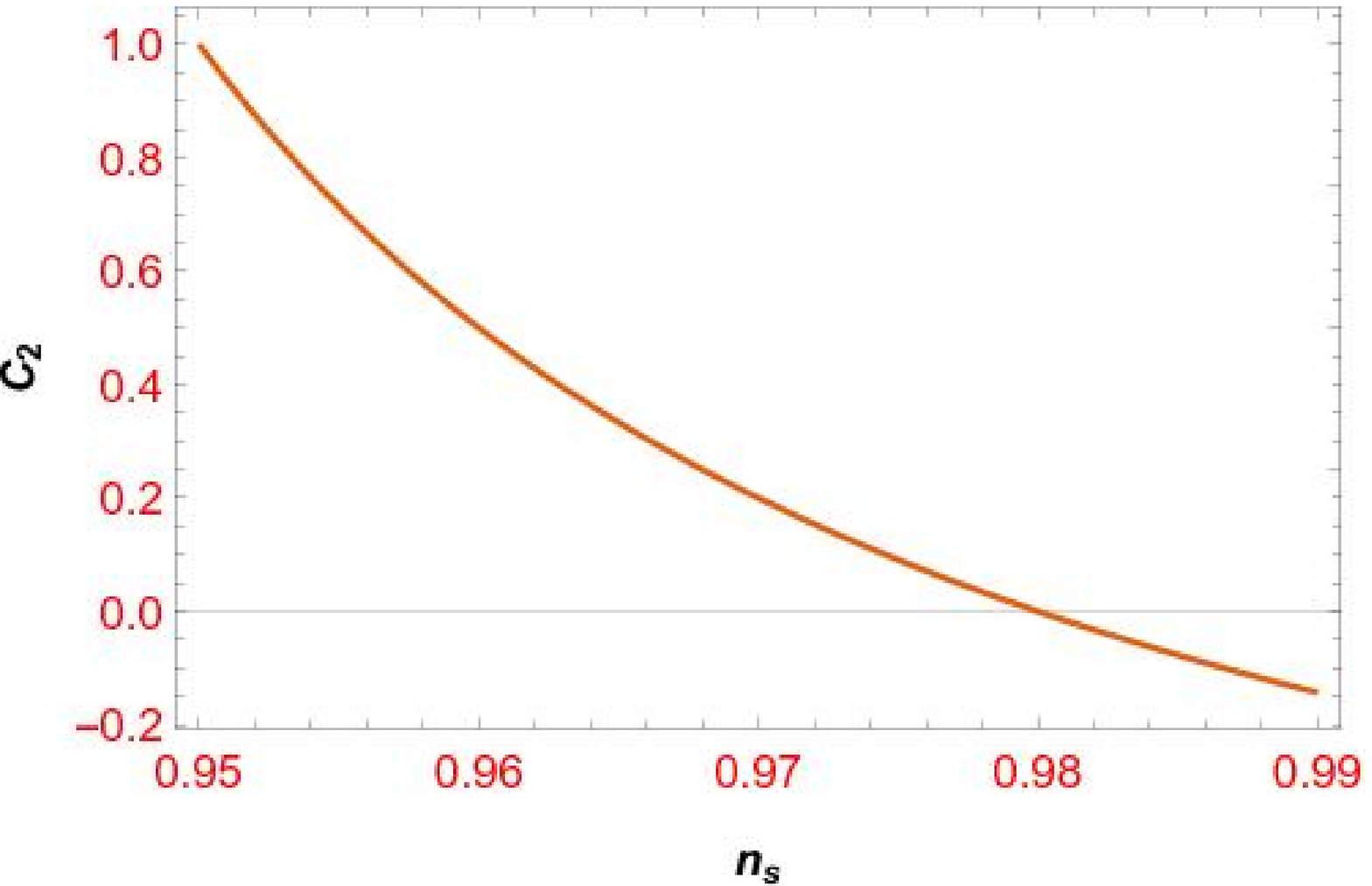}
\label{5b}}
\caption{\small{The plot of $C_{1, 2}$ in terms of $n_{s}$ with respect to free parameter values as $(V_{1}=-1.1\times10^{-20}, V_{0}=1.1\times10^{-20}, \omega_{0}=20)$ }}
\label{5}
\end{center}
\end{figure}

\begin{figure}[h!]
\begin{center}
\subfigure[]{
\includegraphics[height=6cm,width=6cm]{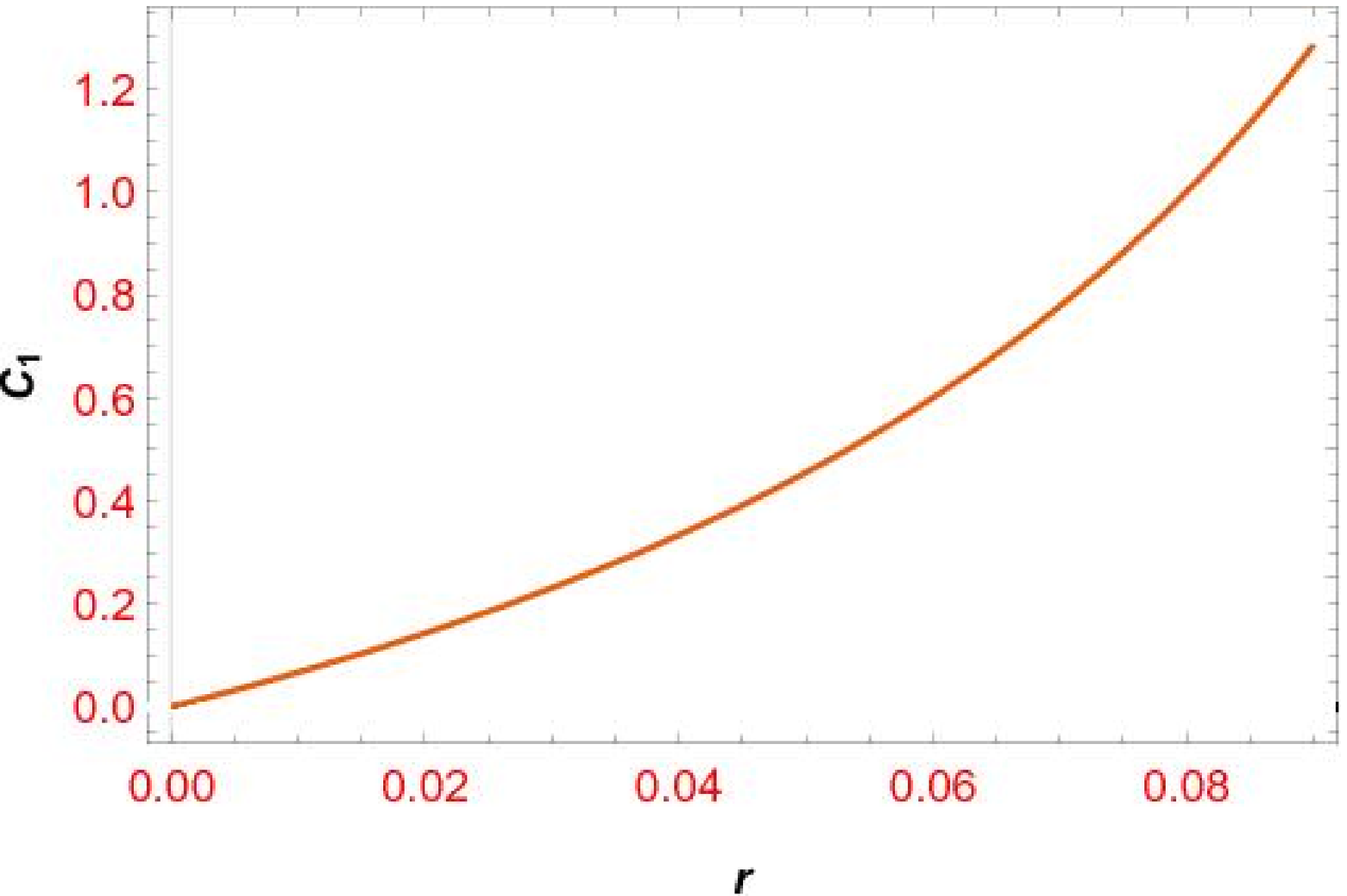}
\label{6a}}
\subfigure[]{
\includegraphics[height=6cm,width=6cm]{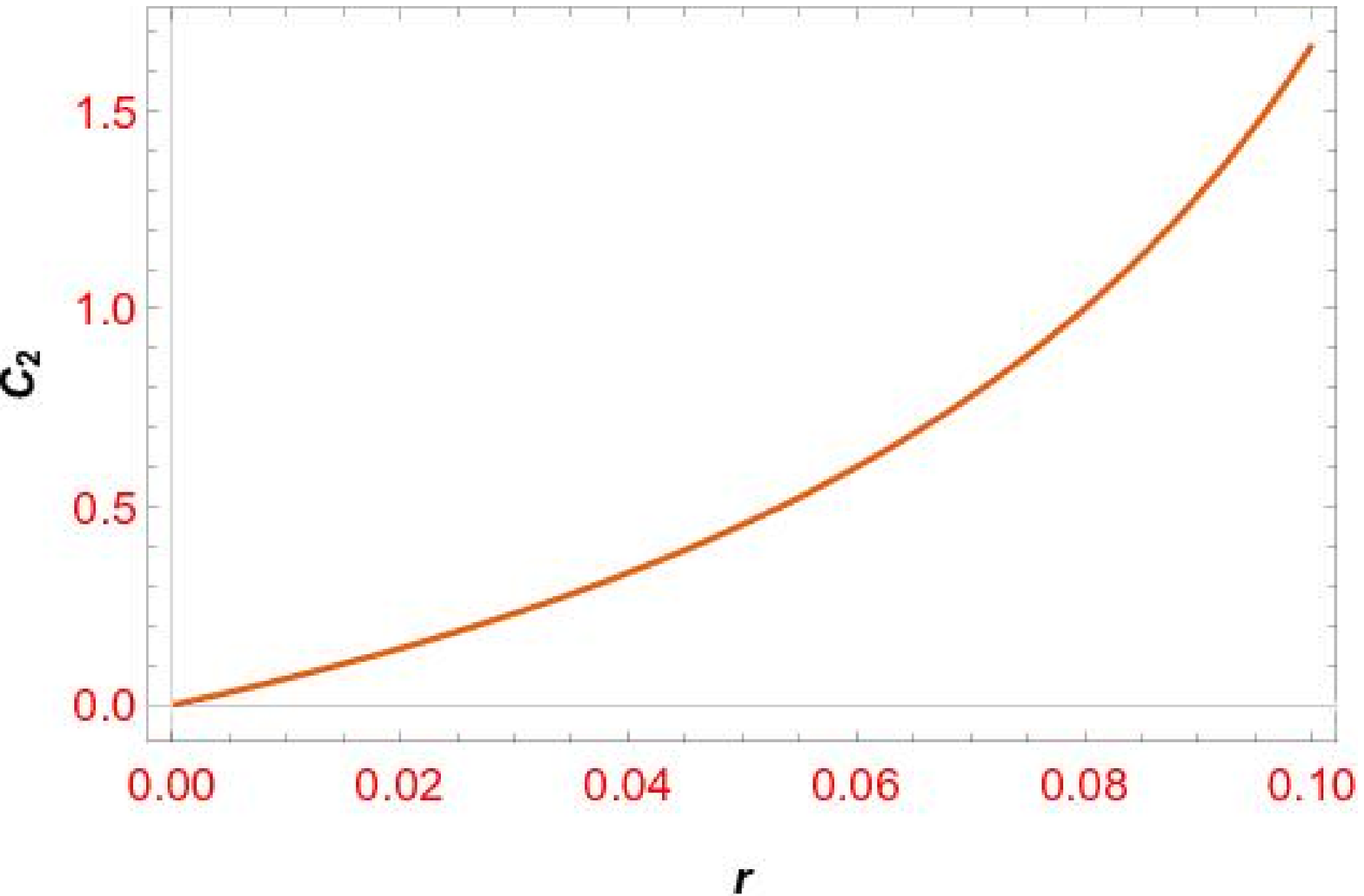}
\label{6b}}
\caption{\small{The plot of $C_{1, 2}$ in terms of $r$ with respect to free parameter values as $(V_{1}=-1.1\times10^{-20}, V_{0}=1.1\times10^{-20}, \omega_{0}=20)$ }}
\label{6}
\end{center}
\end{figure}

\begin{figure}[h!]
 \begin{center}
 \subfigure[]{
 \includegraphics[height=6cm,width=6cm]{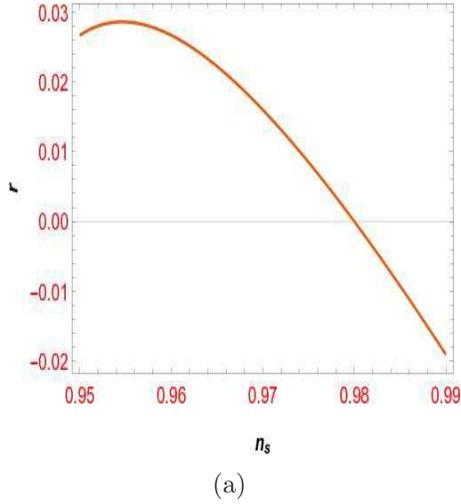}
 \label{7}}
 \caption{The plot of $r$ in term of $n_{s}$ with respect to the free parameter values $(V_{1}=-1.1\times10^{-20}, V_{0}=1.1\times10^{-20}, \omega_{0}=20)$}
 \label{7}
 \end{center}
 \end{figure}

\begin{figure}[h!]
 \begin{center}
 \subfigure[]{
 \includegraphics[height=6cm,width=6cm]{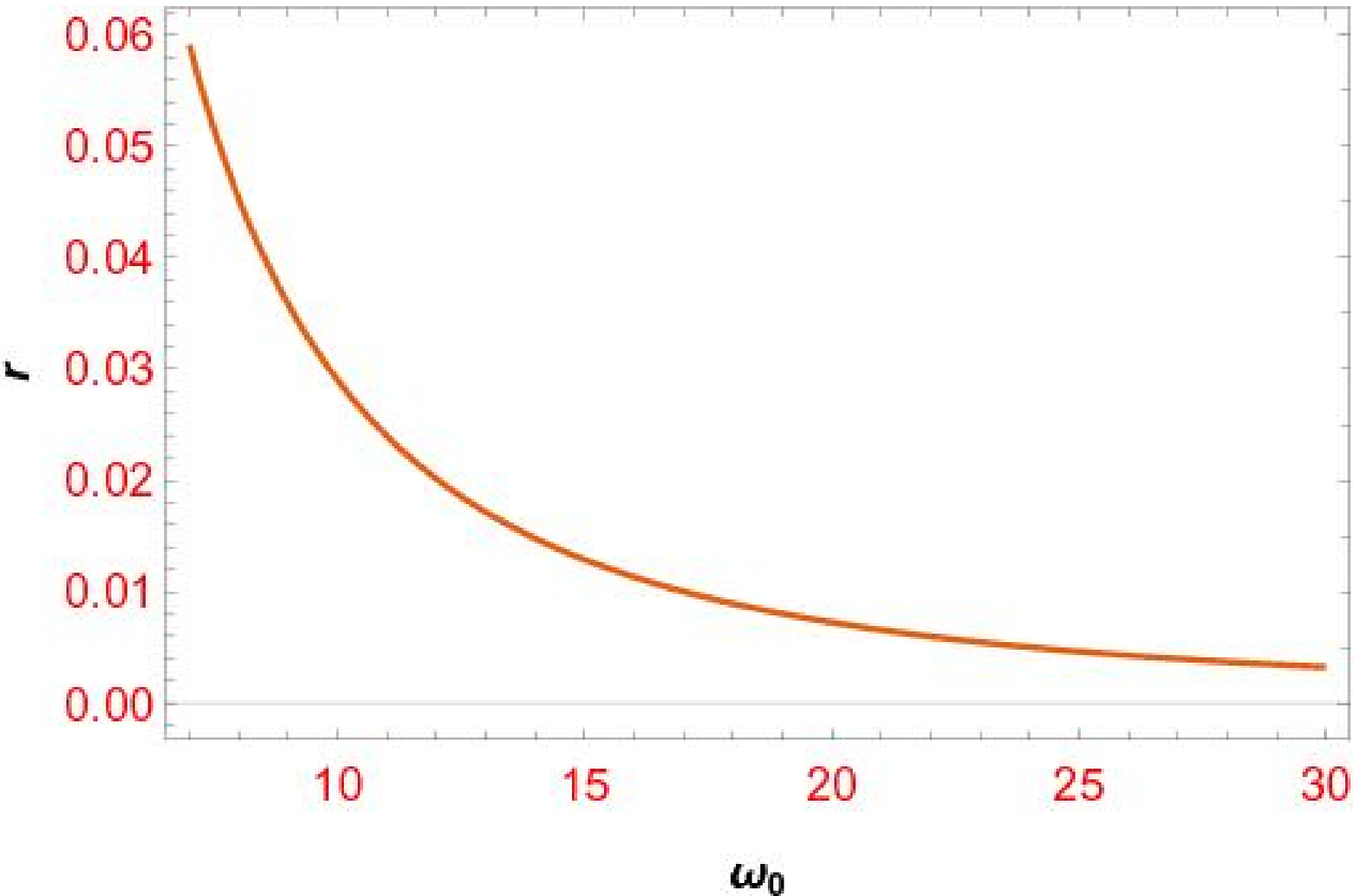}
 \label{8}}
 \caption{The plot of $r$ in term of $\omega_{0}$ with respect to the free parameter values as $(V_{1}=-1.1\times10^{-20}, V_{0}=1.1\times10^{-20}$}
 \label{8}
 \end{center}
 \end{figure}

In this part,  we perform the calculations for exponential potential. We have plotted some figures about constraints related to swampland conjectures, $n_{s}$ and $r$ like the previous section. The results are following: As shown in figure (5), the constraints of the swampland conjecture, i.e., $C_{1, 2}$ are plotted in terms of the scalar spectral index concerning the free parameters as $(V_{0}, V_{1}, \omega_{0})$. The obtained range
associated with the scalar spectral index is consistent with Planck's observable data\cite{7}. Also, for the swampland conjecture coefficient, the range of these components is well defined. Also, as shown in figure (5), the coefficient of $C_{2}$ is smaller than $C_{1}$. Similarly, in figure (6), the constraint of swampland components to the tensor-to-scalar ratio is specified. As shown in figure (6), the range associated with $r$ and the swampland conjectures coefficient are also well defined\cite{7}. To plot these figures, we also used free parameters. In figure(7), we show the restrictions of two cosmological parameters with respect to each other, and the allowable range of these two parameters is also well shown. As observed in the calculations and diagrams of this section, the model of exponential potential in combination with the scalar-tensor theory of gravity and swampland conjectures is consistent with the observable data. Also, in figure (8), we show the limitations $r$ in terms of $\omega_{0}$. As shown in figure (8), the variation of tensor-to-scalar ratio for different values of the parameter $\omega_{0}$ is specified. Like the previous subsection for the corresponding model, the allowable limit of $C_{i}$ can be specified by figure(9).
\begin{figure}[h!]
 \begin{center}
 \subfigure[]{
 \includegraphics[height=6cm,width=6cm]{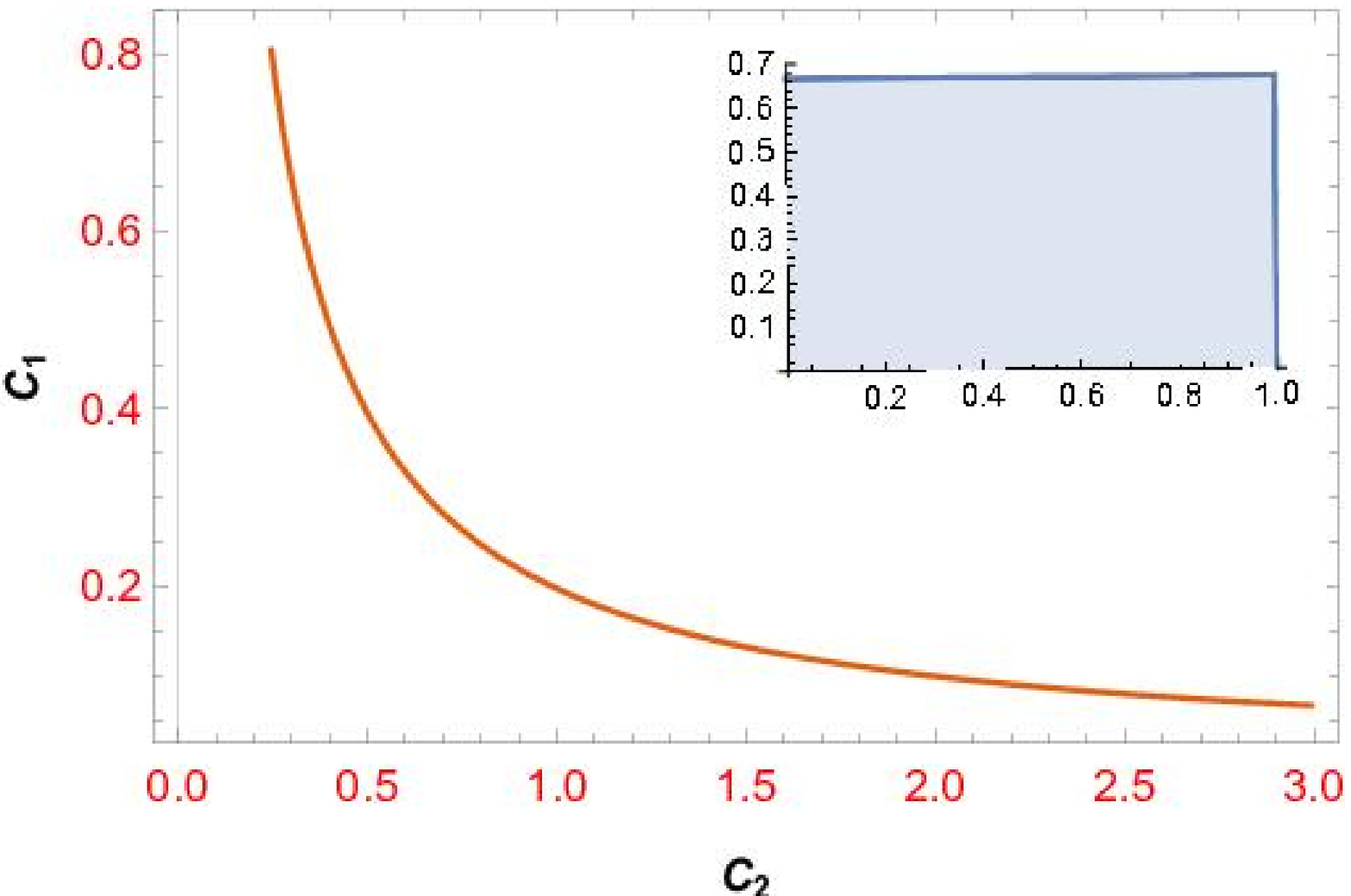}
 \label{9}}
 \caption{The constraints for $C_{1}$, and $C{2}$ with respect to the free parameter values as $(V_{1}=-1.1\times10^{-20},  V_{0}=1.1\times10^{-20}, \omega_{0}=20)$ and tensor-to-scalar ratio $r$ and scalar spectral index $n_{s}$}
 \label{9}
 \end{center}
 \end{figure}

\subsection{Model III; logarithmic Potential}

Here also we consider a logarithmic potential model. So by choosing the
$f(\phi)=(\log(\phi))^{\frac{n}{2}}$,
the potential is $V(\phi)=(\log(\phi))^{n}$. Therefore we will have.

\begin{equation}\label{44}
V(\phi)=V_{1}(\log(\phi))^{n}+V_{0},
\end{equation}
according to equations (12), (16), (17),  (20), (21) and (22), one can obtain,
\begin{equation}\label{45}
\omega(\phi)=\frac{(\omega_{0}^{2}-\frac{3n^{2}(\log(\phi))^{-2+n}}{4\phi^{2}})
\times \phi}{2(\log(\phi))^{\frac{n}{2}}},
\end{equation}
 \begin{equation}\label{46}
V_{E}=V_{1}+V_{0}(\log(\phi))^{-n},
\end{equation}

and the $\frac{d\sigma}{d\phi}$ is,

\begin{equation}\label{47}
\frac{d^{2}\sigma}{d\phi^{2}}=\frac{\omega_{0}^{2}}{(\log(\phi))^{-n}},
\end{equation}

Also here, the slow-roll parameters are given by,
\begin{equation}\label{48}
\epsilon=\frac{V_{0}^{2}n^{2}(\log(\phi))^{n-2}}{\omega_{0}^{2}\phi^{2}(V_{0}+V_{1}(\log(\phi))^{n})^{2}},
\end{equation}
 and
\begin{equation}\label{49}
\eta=\frac{V_{0}n(\log(\phi))^{-2+n}(2+n+2(\log(\phi)))}{\omega_{0}^{2}\phi^{2}(V_{0}+V_{1}(\log(\phi))^{n})},
\end{equation}

We obtain the first and second derivatives of the effective potential
according to equation (46). So we will have,

\begin{equation}\label{50}
V'(\phi)=-\frac{V_{0}n(\log(\phi))^{-1-n}}{\phi}, \hspace{12pt}V''(\phi)=\frac{V_{0}n(\log(\phi))^{-2-n}(1+n+\log(\phi))}{\phi^{2}},
\end{equation}

Now we use the equations (23), (46),  (50) and calculate the following expressions,
\begin{equation}\label{51}
\frac{-\frac{V_{0}n(\log(\phi))^{-1-n}}{\phi}}{V_{1}+V_{0}(\log(\phi))^{-n}}=>C_{1},
\hspace{12pt}\frac{\frac{V_{0}n(\log(\phi))^{-2-n}(1+n+\log(\phi))}{\phi^{2}}}{V_{1}+
V_{0}(\log(\phi))^{-n}}<-C_{2},
\end{equation}

considering the equations (48) and (49), $r$ and $n_{s}$, one can obtain.

\begin{equation}\label{52}
n_{s}=1-\frac{6V_{0}^{2}n^{2}(\log(\phi))^{n-2}}{\omega_{0}^{2}\phi^{2}(V_{0}+V_{1}(\log(\phi))^{n})^{2}}+\frac{2V_{0}n(\log(\phi))^{-2+n}(2+n+2(\log(\phi)))}{\omega_{0}^{2}\phi^{2}(V_{0}+V_{1}(\log(\phi))^{n})},
\end{equation}
 and
\begin{equation}\label{53}
r=\frac{16V_{0}^{2}n^{2}(\log(\phi))^{n-2}}{\omega_{0}^{2}\phi^{2}(V_{0}+V_{1}(\log(\phi))^{n})^{2}},
\end{equation}

Now by considering the mentioned constant values for the free parameters, the scalar spectral index and tensor-to-scalar ratio for $n=3, 4 ,5$ will be $n_{s}=0.984, 0.976, 0.963$ and $r=0.066, 0.053, 0.038$, respectively. According to the component $f(\phi)$ also $\phi_{i}=1.5$, $V_{0}=0.9\times 10^{-13}T^{-2}$ and $V_{1}=0.9\times 10^{-13}T^{-2}$ it can be checked that different values obtained for the parameters under consideration. Hence we will have $10<\omega_{0}<30$, $0.7124<\phi_{e}<1.4315$, $0.0.011<r<0.0.084$, $0.9523<n_{s}<0.9892$ and $42<N<61$.  Also this rout can be studied for different values of the parameter $n$ according to the $f(\phi)$. Now, like the pervious section, we reverse these equations (52) and (53) in terms of two parameters $n_{s}$ and $r$ and then placed in equation (51). Then we will plot the figures $C_{1, 2}-n_{s}$ and $C_{1, 2}-r$, and $r-n_{s}$ respectively. So, according to the above description we have some figures,

\begin{figure}[h!]
\begin{center}
\subfigure[]{
\includegraphics[height=6cm,width=6cm]{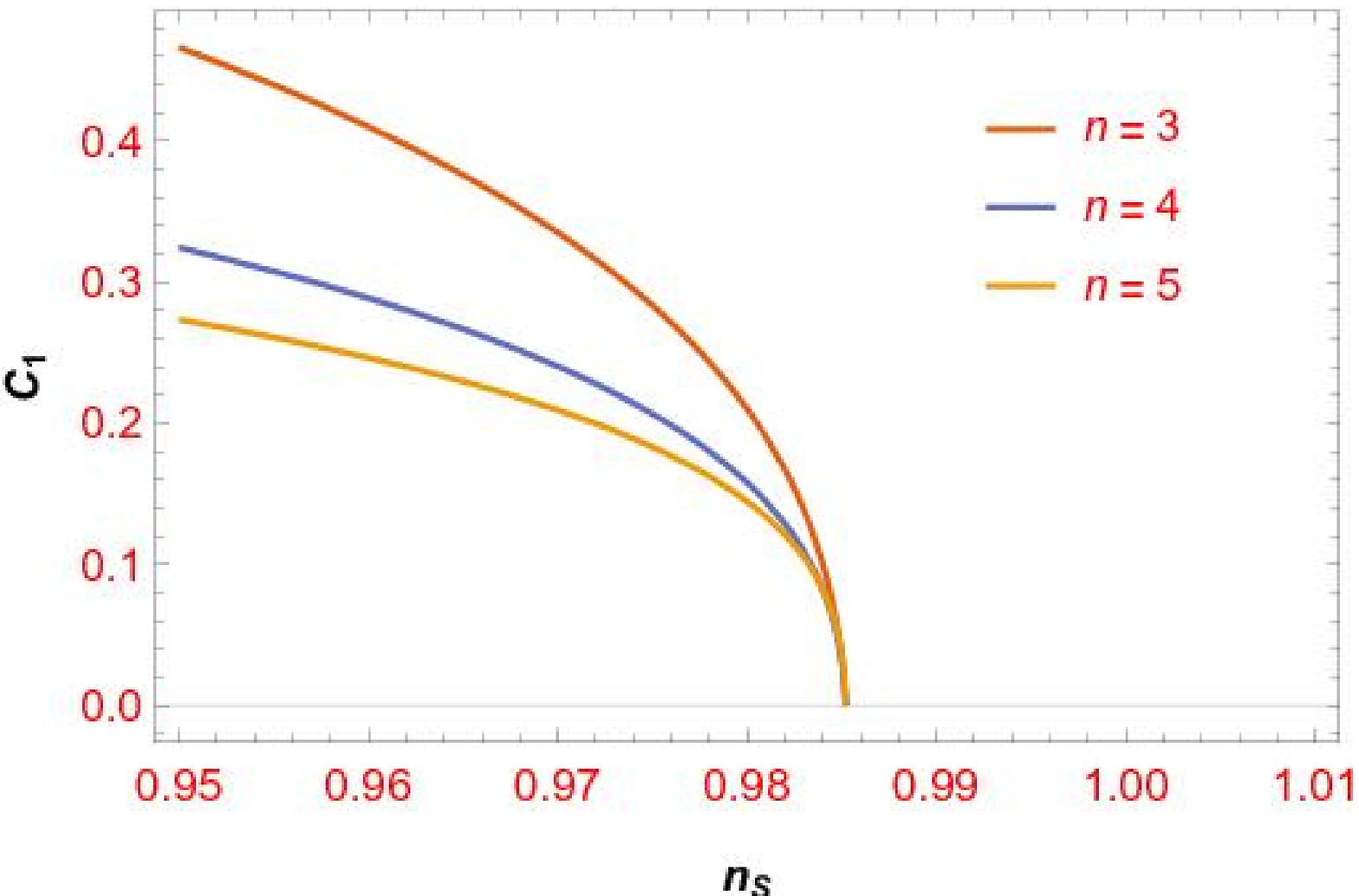}
\label{10a}}
\subfigure[]{
\includegraphics[height=6cm,width=6cm]{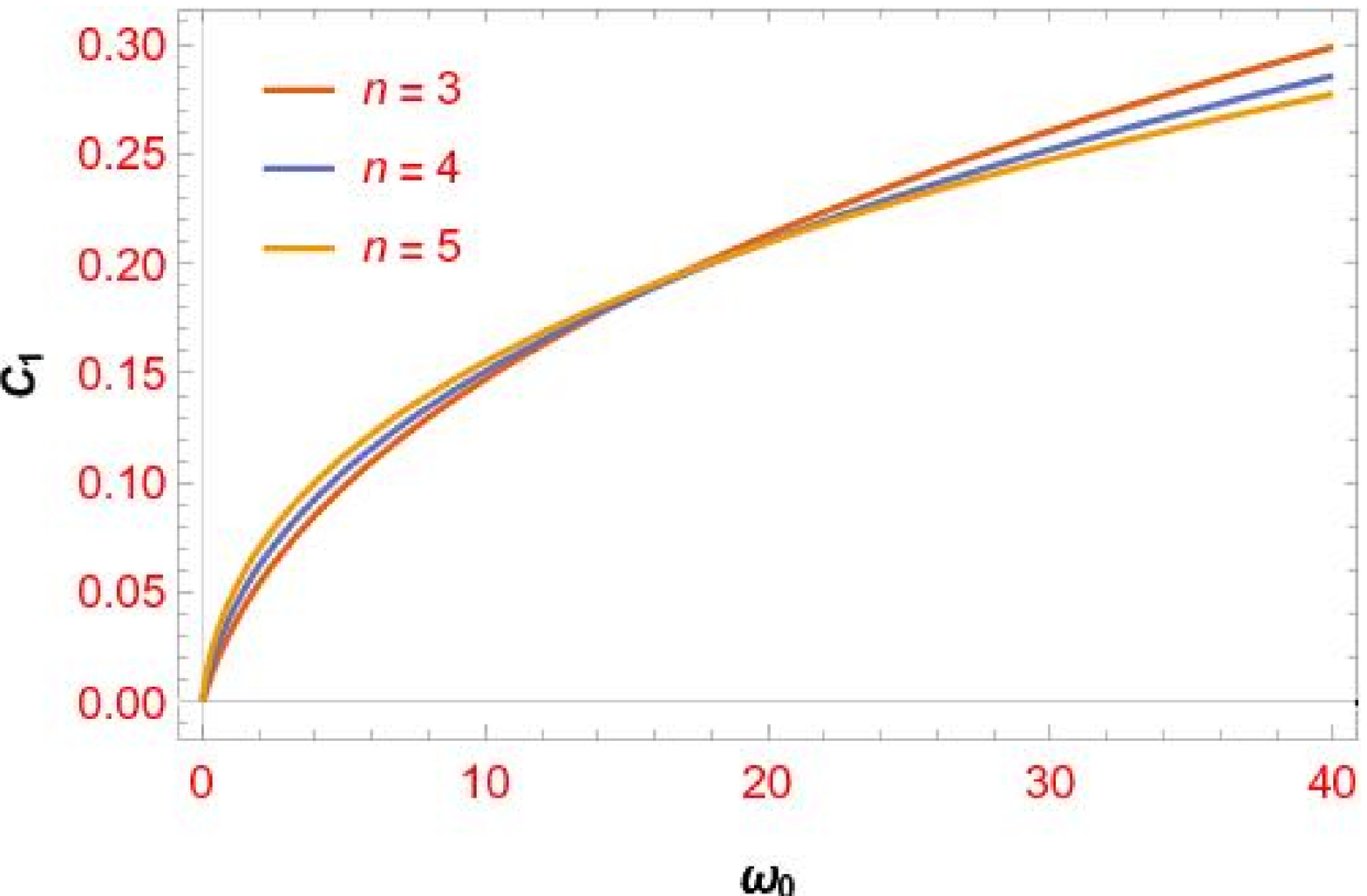}
\label{10b}}
\caption{\small{The plot of $C_{1}$ in terms of $n_{s}$ and $\omega_{0}$ with respect to free parameter values as $(V_{1}=-0.9\times10^{-13}, V_{0}=0.9\times10^{-13}, \omega_{0}=20)$ }}
\label{10}
\end{center}
\end{figure}

\begin{figure}[h!]
\begin{center}
\subfigure[]{
\includegraphics[height=6cm,width=6cm]{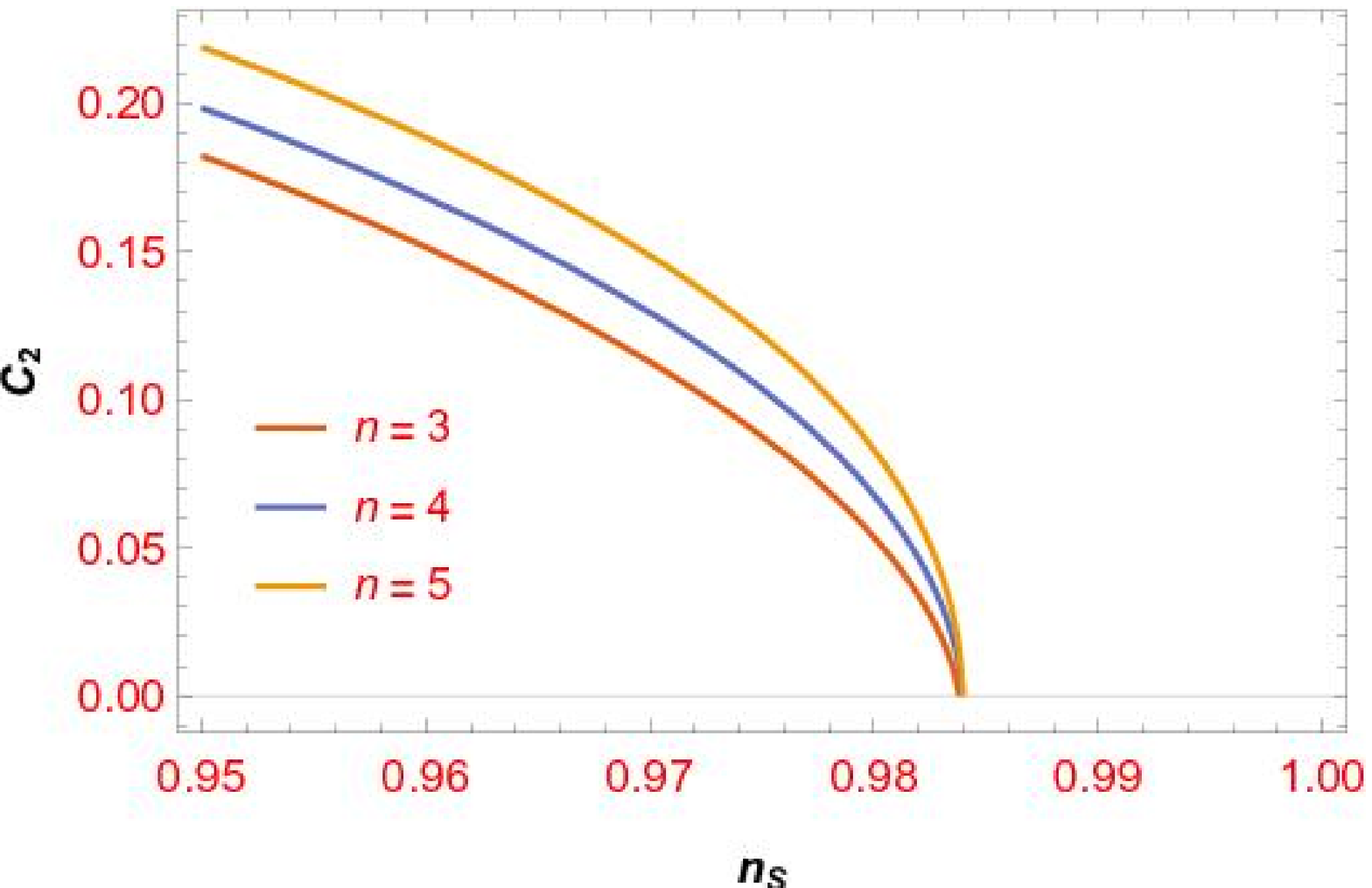}
\label{11a}}
\subfigure[]{
\includegraphics[height=6cm,width=6cm]{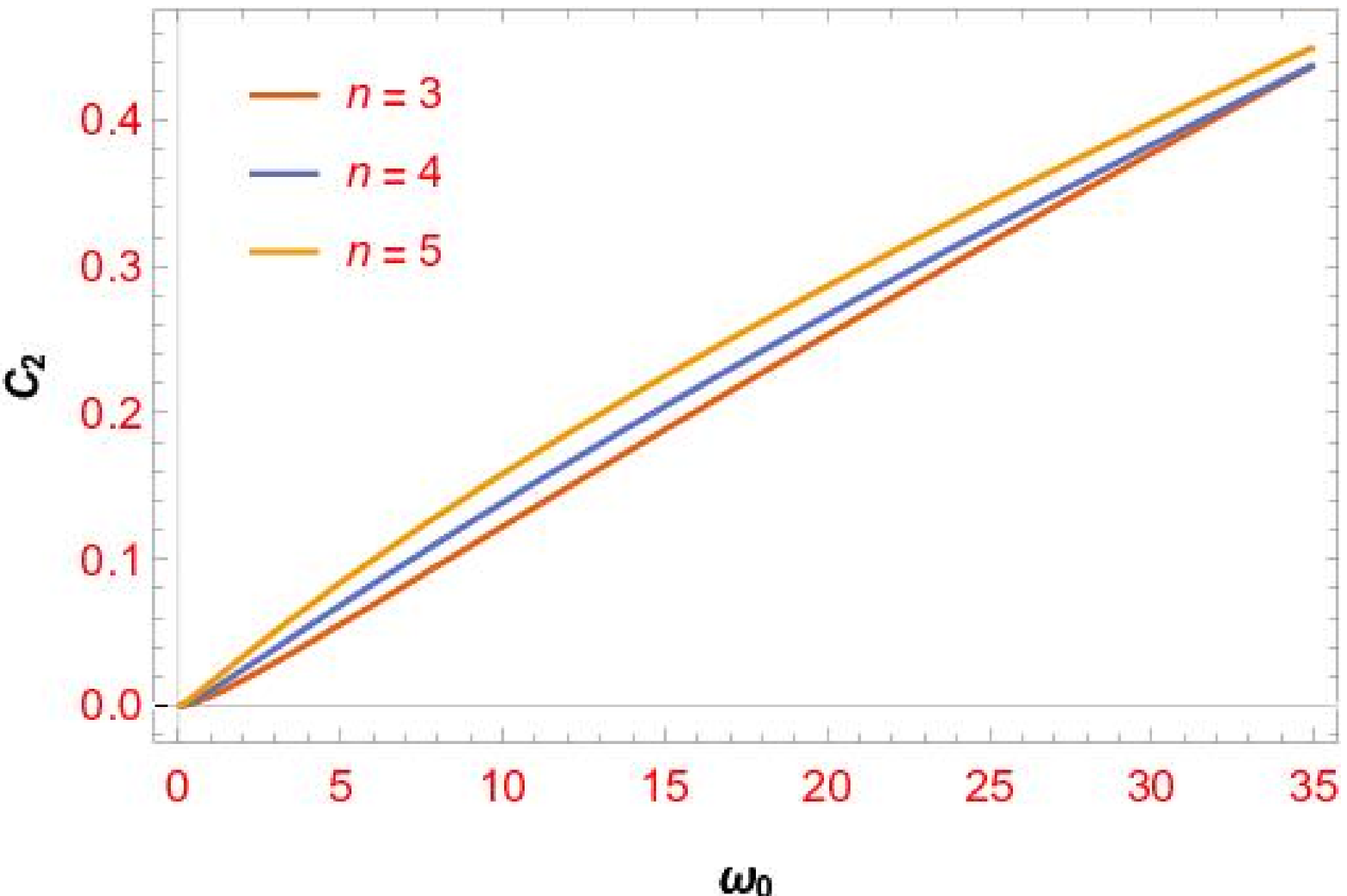}
\label{11b}}
\caption{\small{The plot of $C_{2}$ in terms of $n_{s}$ and $\omega_{0}$ with respect to free parameter values as $(V_{1}=-0.9\times10^{-13}, V_{0}=0.9\times10^{-13}, \omega_{0}=20)$ }}
\label{11}
\end{center}
\end{figure}

\begin{figure}[h!]
\begin{center}
\subfigure[]{
\includegraphics[height=6cm,width=6cm]{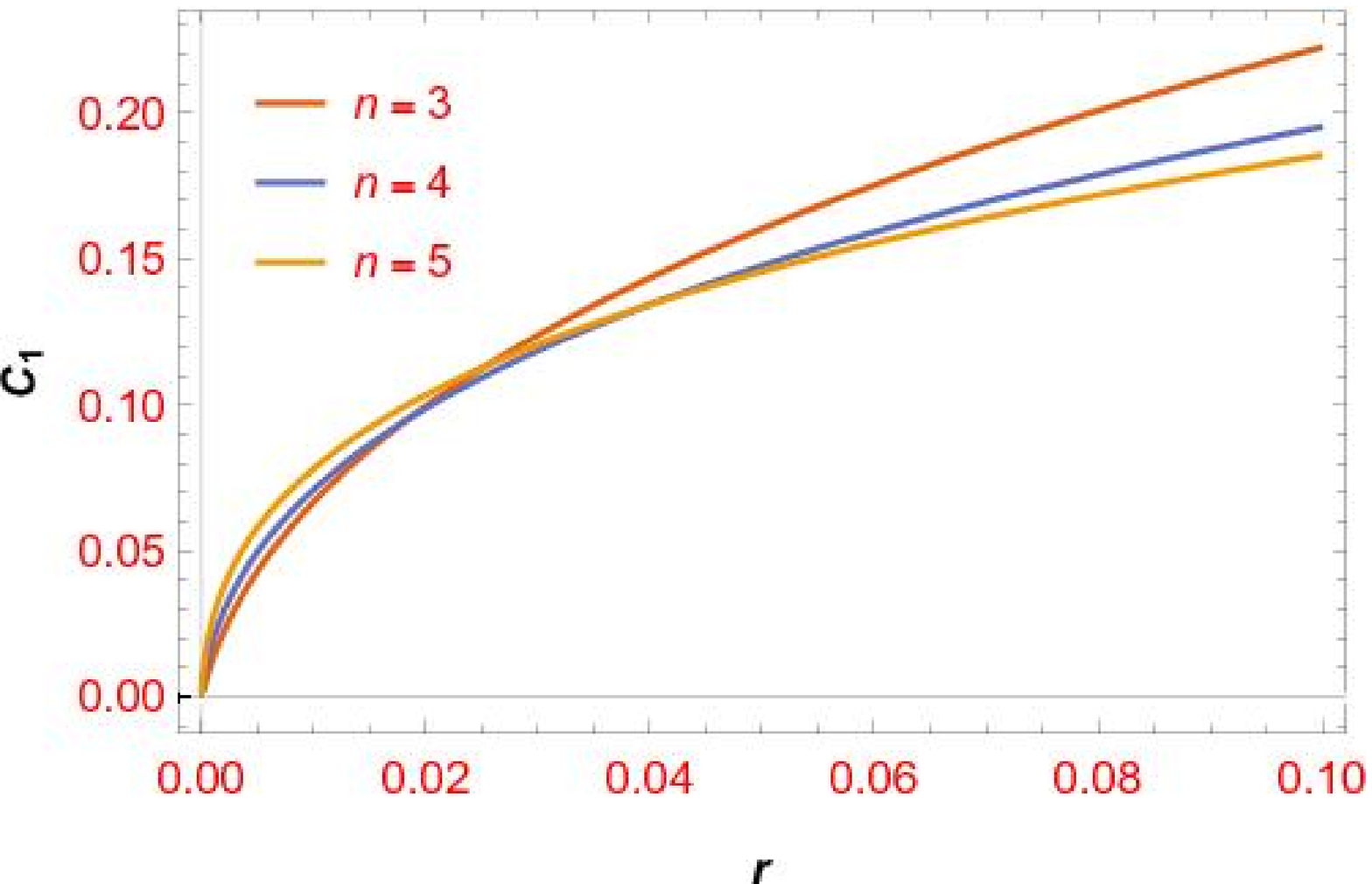}
\label{12a}}
\subfigure[]{
\includegraphics[height=6cm,width=6cm]{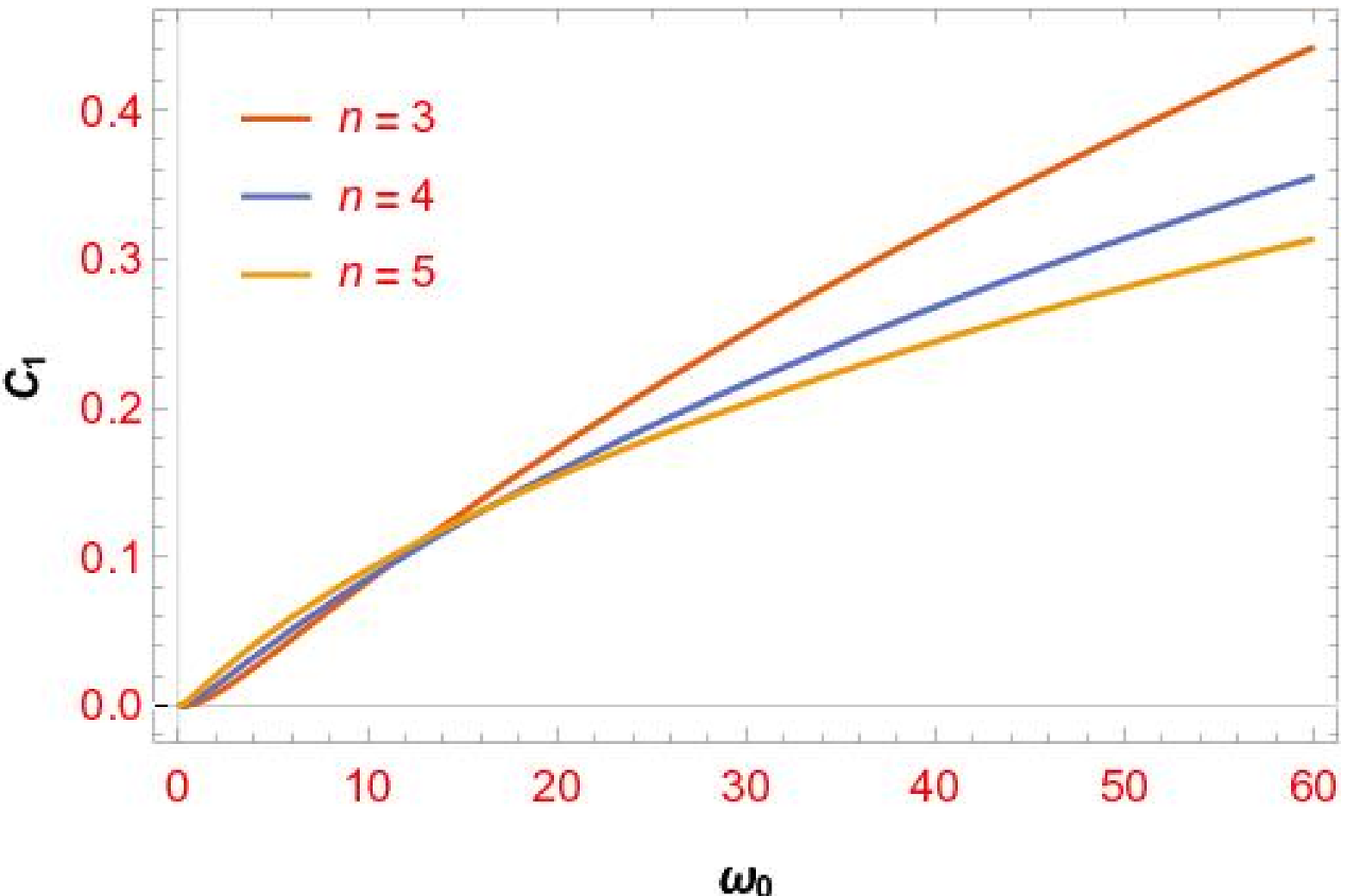}
\label{12b}}
\caption{\small{The plot of $C_{1}$ in terms of $r$ and $\omega_{0}$ with respect to free parameter values as $(V_{1}=-0.9\times10^{-13}, V_{0}=0.9\times10^{-13}, \omega_{0}=20)$ }}
\label{12}
\end{center}
\end{figure}

\begin{figure}[h!]
\begin{center}
\subfigure[]{
\includegraphics[height=6cm,width=6cm]{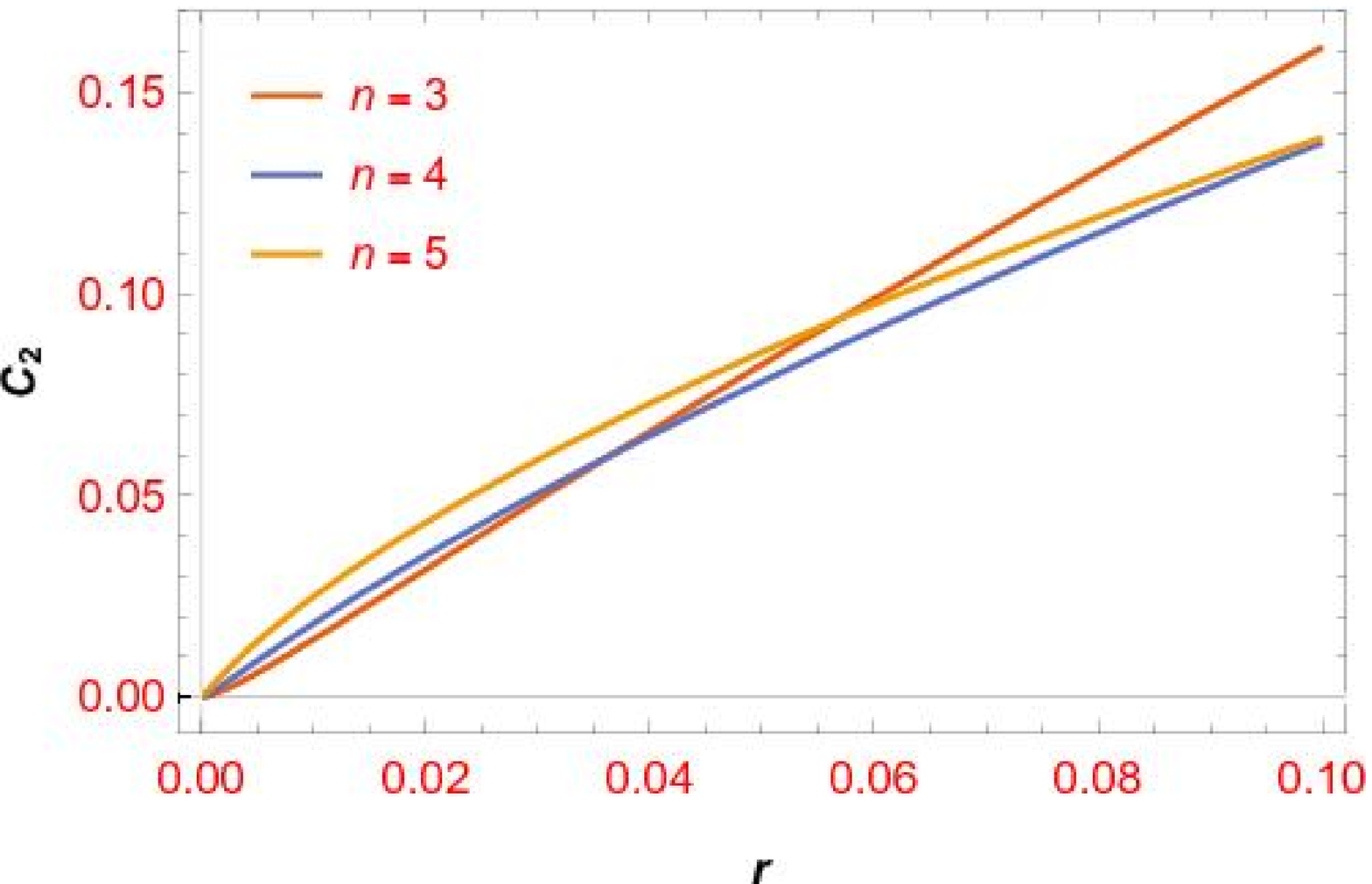}
\label{13a}}
\subfigure[]{
\includegraphics[height=6cm,width=6cm]{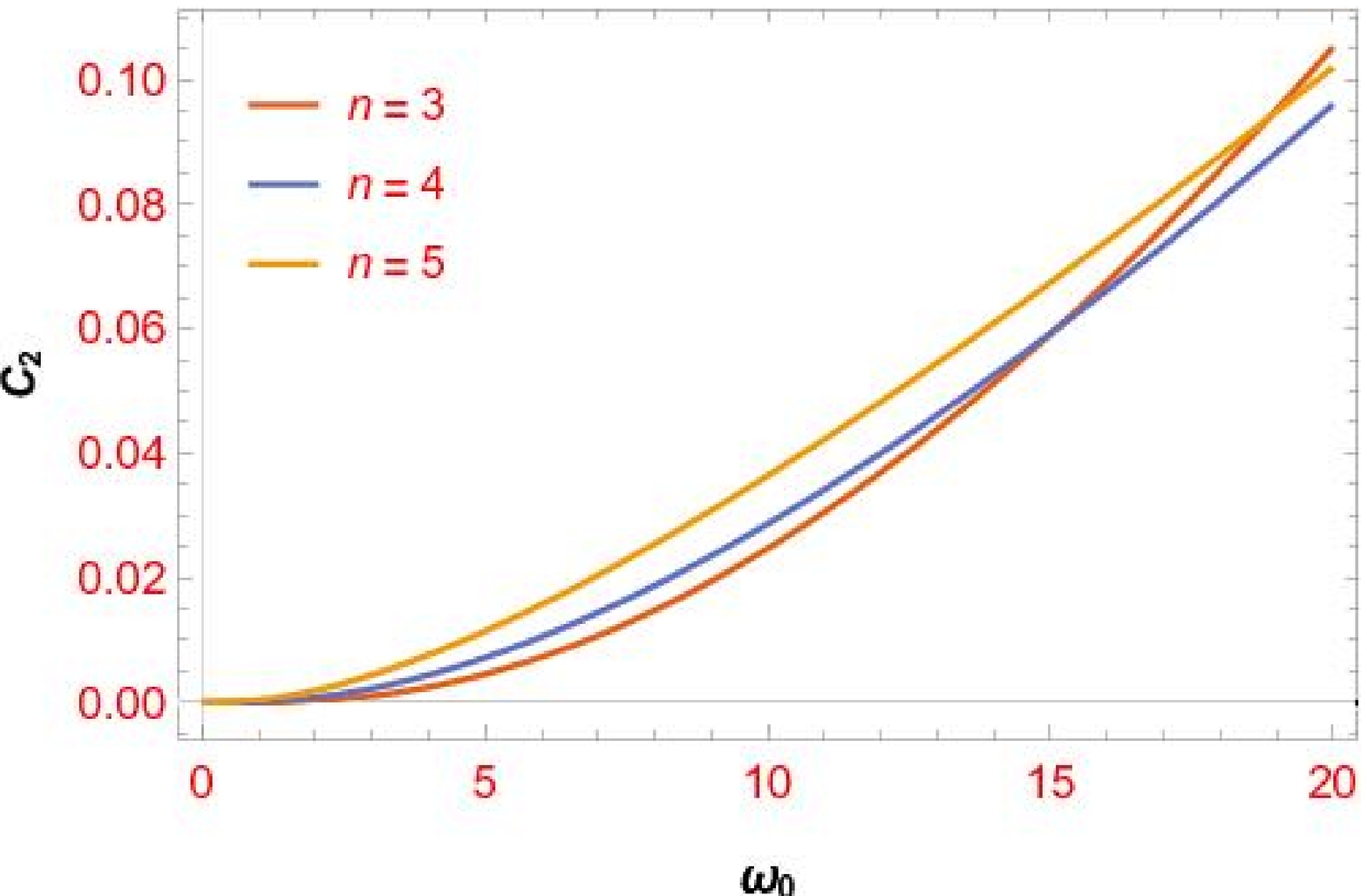}
\label{13b}}
\caption{\small{The plot of $C_{2}$ in terms of $r$ and $\omega_{0}$ with respect to free parameter values as $(V_{1}=-0.9\times10^{-13}, V_{0}=0.9\times10^{-13}, \omega_{0}=20)$ }}
\label{13}
\end{center}
\end{figure}

\begin{figure}[h!]
 \begin{center}
 \subfigure[]{
 \includegraphics[height=6cm,width=6cm]{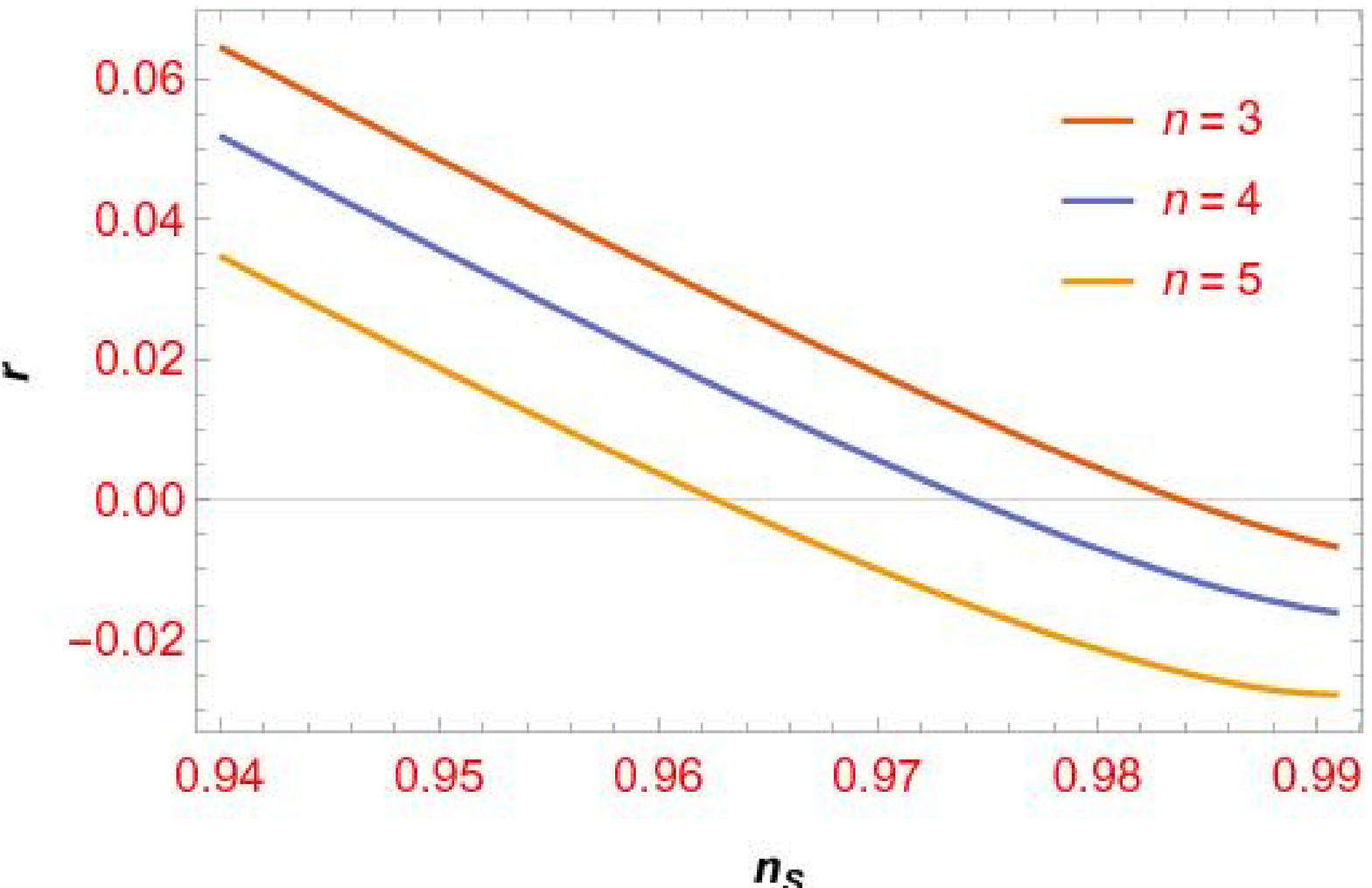}
 \label{14}}
 \caption{The plot of $r$ in term of $n_{s}$ with respect to the free parameter values as $(V_{1}=-0.9\times10^{-13}, V_{0}=0.9\times10^{-13}, \omega_{0}=20)$}
 \label{14}
 \end{center}
 \end{figure}

\begin{figure}[h!]
 \begin{center}
 \subfigure[]{
 \includegraphics[height=6cm,width=6cm]{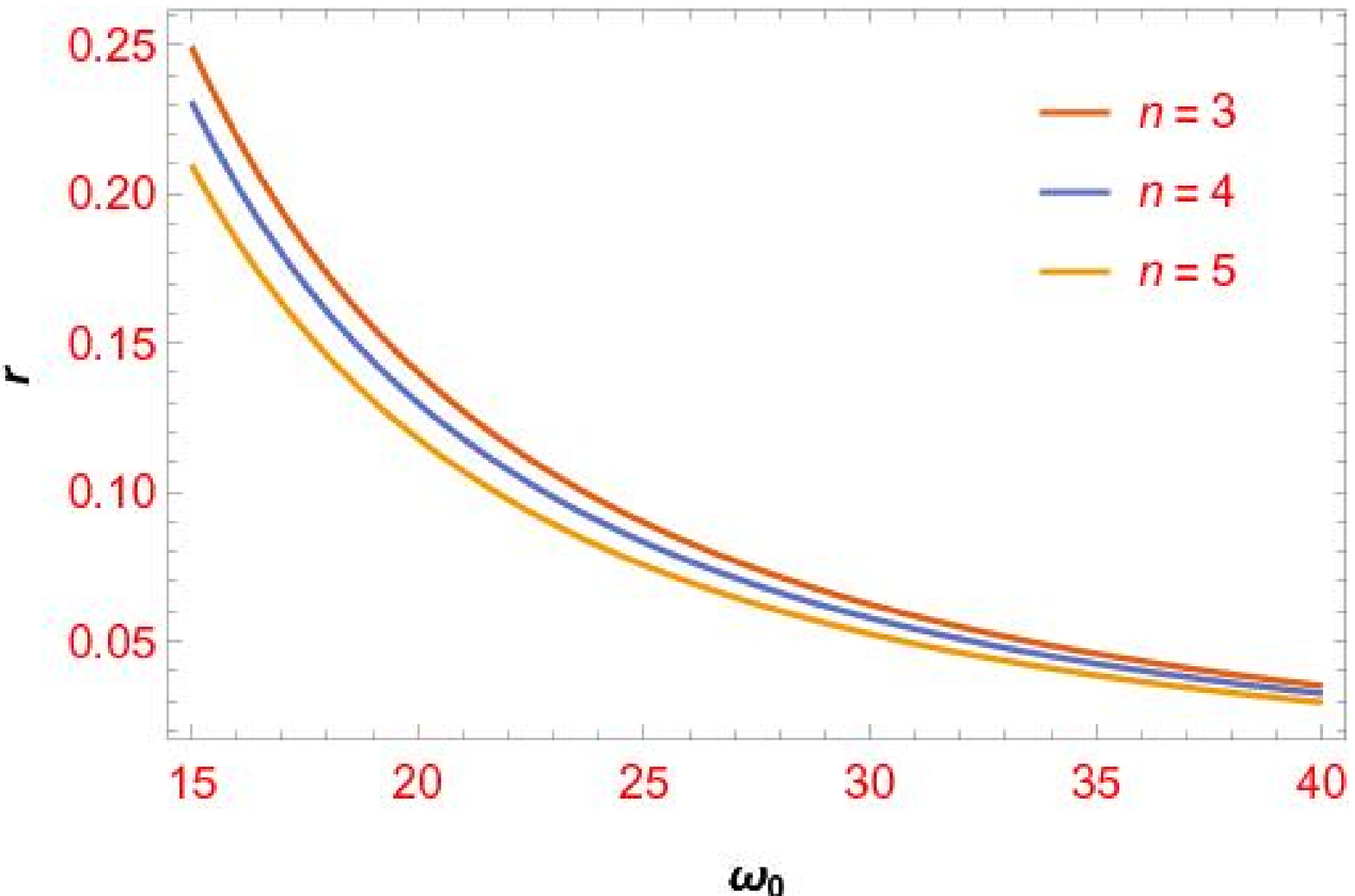}
 \label{15}}
 \caption{The plot of $r$ in term of $\omega_{0}$ with respect to the free parameter values as $(V_{1}=-0.9\times10^{-13}, V_{0}=0.9\times10^{-13}$}
 \label{15}
 \end{center}
 \end{figure}
As two previous sections, we consider the same limitations for the logarithmic model. Of course, in this section, we also specify the role of the parameter $n$. As mentioned above, we have no acceptable answers for $(n<3)$, but for the $(n\geq3)$, the result was comparable to the observable data\cite{7}. So we examined the extent of these constraints by considering the three different values of $n$ viz the swampland conjectures in terms of two cosmological parameters $n_{s}$ and $r$. In figure (10), the constraints of the $C_{1}$ are plotted in terms of the scalar spectral index and the $\omega_{0}$. The range of scalar spectral index concerning different values of $n$ as 3,4,5 is consistent with Planck's observable data\cite{7}. Also, the changes of this component in terms of the parameter $\omega_{0}$ for three different values of $n$ are well specified.  The range of the swampland conjecture components is well defined, which is usually specified by the literature as order 0.1 and 1. As shown in figures, coefficient of $C_{1}$ for larger values of the parameter $\omega_{0}$, it assigns smaller values. Similarly, in figure (11), constraints of $C_{2}$ are examined in terms of the scalar spectral index and the $\omega_{0}$. As shown in figure (11), the range of  $n_{s}$, $\omega_{0}$ and swampland conjectures coefficient are also well defined for various values of $n$. In this figure, unlike the previous section, for larger values of $\omega_{0}$, the  $C_{2}$ has larger values. Also, as shown in figure (11a), the allowable values of $C_{2}$ are smaller than $C_{1}$. To plot these figures, we also used free parameters. We plotted the restrictions of $C_{1 2}$ in terms of the tensor-to scalar ratio and the $\omega_{0}$ by considering different values of $n$ In figures (12) and (13). Also, in figure (12b), for larger values of $n$, the values of $C_{1}$ become smaller. But in figure (13b), for larger values of $n$, the values of $C_{1}$ become bigger. Also, the values of the component
 $C_{2}$ in figure (13a) have smaller values than the component $C_{1}$ in figure (12a). As shown in figure (14), we plot two cosmological parameters with respect to each other. The allowable range of these two parameters is well shown in this figure for different $n$. As shown in figure (14), for larger values of $n$, the result will be closer to the observable data\cite{7}. Also, in figure (15), we plotted the change of this cosmological parameter, i.e., $r$, in terms of different values of $\omega_{0}$. The large values of $\omega_{0}$ are more acceptable than the values for $r$ so that they are comparable to the observable data. In these two figures, we also used the free parameters. We note here; these variations are displayed for different values of $n$. According to the calculations and figures plotted in this part, the logarithmic model in combination with the scalar-tensor theory of gravity and swampland conjectures is consistent with the observable data\cite{7} This compatibility was more seen for larger values $n$ and $\omega_{0}$. Scalar-tensor theory of gravity can also be well compatible with swampland conjectures. For this model, the allowable range of swampland conjectures $C_{i}$ for different values of $n$ is determined according to the constant parameters as well as the observable data $r$ and $n_{s}$. In figure (16), we see the allowable ranges of swampland components or $C_{i}$ for the corresponding model.

\begin{figure}[h!]
 \begin{center}
 \subfigure[]{
 \includegraphics[height=6cm,width=6cm]{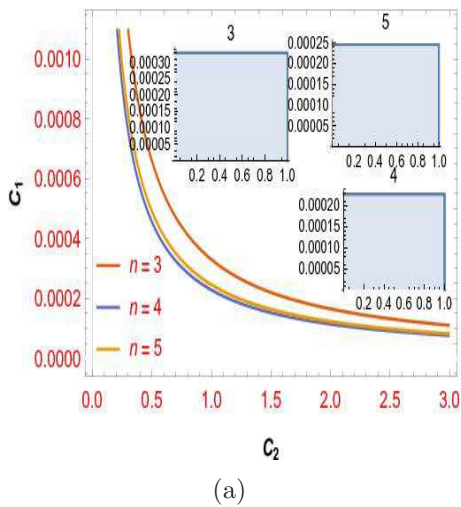}
 \label{16}}
 \caption{The constraints for $C_{1}$, and $C{2}$ with respect to the free parameter values as $(V_{1}=-0.9\times10^{-13},  V_{0}=0.9\times10^{-13}, \omega_{0}=13 )$, $n=3, 4, 5$ and tensor-to-scalar ratio $r$ and scalar spectral index $n_{s}$}
 \label{16}
 \end{center}
 \end{figure}

It is clear that from studying all three models, the logarithmic model with comparing to the other models with increasing the values of $n$, the cosmological parameters $r$ and $n_{s}$ more compatible with observable data. After that, the exponential model and finally the power-law model is more appropriate. The constraints of the $C_{1}$ and $C_{2}$ for these three models are shown in figures (4), (9), and (16) viz they are satisfied by swampland conditions. Here also, we note that $C_{i}s$ permitted for each model in a certain range. The range of these components, i.e., $C_{1, 2}$ is calculated positive unit order and in some calculations is considered approximately $0.1$ in the literature. As you can see in figures (4), (9), and (16), the exponential model is within the computational range of the literature. However,  we can not ignore the other two models; that is, swampland conjectures can somehow accommodate a larger volume of the exponential model and then the power-law and finally logarithmic.
\section{Conclusion}
Cosmic inflation with various theories studied from different perspectives and conditions. In this paper, we checked inflation from the scalar-tensor gravity perspective and swampland conjectures. Therefore, by selecting different inflation models as power-law,  exponential and logarithmic, we calculated potential, scalar spectral index and tensor-to-scalar ratio in the framework of scalar-tensor gravity. Then we employed the swampland conjectures on the  scalar-tensor gravity in inflation and  examined new constraints such as $C_{1,2}-n_{s}$, $C_{1,2}-r$ and $r-n_{s}$. We also determined the range of these cosmological parameters and swampland conjectures coefficient and compared these results to the latest observable data such as Planck 2018.  It has imposed new tighter constraints on cosmological parameters. Finally, we analyzed the obtained results and achieved the compatibility or incompatibility of this inflation model with the swampland conjectures.  As observed in the calculations and figures of this paper, we showed that the logarithmic model with comparing to the other models with increasing the values of $n$, the cosmological parameters $r$ and $n_{s}$ more compatible with observable data. But from equations of  (4), (9), and (16) and corresponding figures, we observed that the exponential model is within the computational range of the literature. However, we can not ignore the other two models. But swampland conjectures are more compatible first to the exponential model and then the power-law and finally logarithmic. So, generally, one can say that the power law, exponential and logarithmic potential in combination with the scalar-tensor theory of gravity and swampland conjectures are consistent with the observable data\cite{7}. Therefore, the scalar-tensor theory of gravity can also be well compatible with swampland conjectures. It may be interesting to check the above corresponding model for the constant-roll with other conditions of the swampland program such as TCC. Also, it can challenge these concepts for other cosmological structures and theories, such as dark energy and dark matter. We leave the calculations related to these structures and concepts to future work. Also, one critical question that arises through these calculations is whether, assuming the swampland constraints, is eternal inflation allowed for such scalar-tensor models?


\begin{thebibliography}{11}
\bibitem{1}
A. A. Starobinsky, Phys. Lett. B91, 99 (1980).
\bibitem{2}
A. H. Guth, Phys. Rev. D23, 347 (1981).
\bibitem{3}
A. R. Liddle, D. H. Lyth, Cambridge University Press (2000).\\
S. Dodelson, Modern Cosmology, Academic Press (1999).
\bibitem{4}
V. F. Mukhanov, H. A. Feldman and R. H. Brandenberger, Phys. Rept. 215, 203 (1992).\\
D. Langlois, Lect. Notes Phys. 800, 1 (2010).
\bibitem{5}
J. Khoury, B. A. Ovrut, P. J. Steinhardt and N. Turok, Phys. Rev. D 64, 123522 (2001).\\
J. Khoury, B. A. Ovrut, P. J. Steinhardt and N. Turok, Phys. Rev. D 66, 046005 (2002).
\bibitem{6}
H. V. Peiris et al., Astrophys. J. Suppl. 148, 213 (2003).\\
G. Hinshaw et al., Astrophys. J. Suppl. 208, 19 (2013).
\bibitem{7}
P. A. R. Ade et al., Astron. Astrophys. 594, A20 (2016).\\
P. A. R. Ade et al., Phys. Rev. Lett. 116, 031302 (2016).\\
P. A. R. Ade et al., Astron.Astrophys. 571 A22(2014).\\
Y. Akrami, et al., Astron. Astrophys. 641, A10 (2020).\\
N. Aghanim, et al., Astron. Astrophys. 641, A6 (2020).
\bibitem{8}
J. E. Lidsey, A. R. Liddle, E. W. Kolb, E. J. Copeland, T. Barreiro and M. Abney, Rev. Mod. Phys. 69, 373 (1997).\\
E. Elizalde, S. Nojiri, S. D. Odintsov, D. Saez-Gomez and V. Faraoni, Phys. Rev. D 77, 106005 (2008).
\bibitem{9}
K. A. Olive, Phys. Rept. 190, 307 (1990).
\bibitem{10}
D. H. Lyth and A. Riotto, Phys. Rep. 314, 1 (1999).
\bibitem{11}
A. R. Liddle and D.H. Lyth,(Cambridge University Press, (2000).
\bibitem{12}
D. Baumann, Contribution to: TASI 2009, 523-686 arXiv:0907.5424 (2009).
\bibitem{13}
J. Martin, C. Ringeval and V. Vennin, Phys. Dark Univ. 5-6, 75-235 (2014).
\bibitem{14}
J. Martin, C. Ringeval, R. Trotta and V. Vennin, JCAP 1403, 039 (2014).
\bibitem{15}
J. Martin, Astrophys. Space Sci. Proc. 45, 41-134 (2016).
\bibitem{16}
S. Nojiri, and S. D. Odintsov, Phys. Rept. 505, 59 (2011).\\
S. Capozziello, and M. De Laurentis, Phys. Rept. 509, 167 (2011).\\
T. Clifton, P. G. Ferreira, A. Padilla, and C. Skordis, Phys. Rept. 513, 1 (2012).\\
S. Capozziello, M. de Laurentis, and V. Faraoni, Open Astron. J. 3, 49 (2010).\\
S. Nojiri, S. D. Odintsov and V. K. Oikonomou, Phys. Rept. 692 (2017).\\
G. J. Olmo, Int. J. Mod. Phys. D 20, 413 (2011).\\
J. Beltran Jimenez, L. Heisenberg, G. J. Olmo and D. RubieraGarcia, Phys. Rept. 727, 1 (2018).
\bibitem{17}
K. Bamba, S. Nojiri, S. D. Odintsov and D. Saez-Gomez, Phys. Rev. D 90, 124061 (2014).
\bibitem{18}
A. A. Starobinsky, Phys. Lett. 91B, 99 (1980).
\bibitem{19}
A. de la Cruz-Dombriz, E. Elizalde, S. D. Odintsov and D. Saez-Gomez, JCAP 1605, 05, 060 (2016).
\bibitem{20}
G. Cognola, E. Elizalde, S. Nojiri, S. D. Odintsov, L. Sebastiani and S. Zerbini, Phys. Rev. D 77, 046009 (2008).\\
S. Nojiri and S. D. Odintsov, Phys. Rev. D 77, 026007 (2008).\\
G. Cognola, E. Elizalde, S. D. Odintsov, P. Tretyakov and S. Zerbini, Phys. Rev. D 79, 044001 (2009).
\bibitem{21}
S. D. Odintsov, D. Saez-Chillon Gomez and G. S. Sharov, Eur. Phys. J. C 77, no. 12, 862 (2017).
\bibitem{22}
T. Harko, F. S. N. Lobo, S. Nojiri, and S. D. Odintsov, Phys. Rev. D 84, 024020 (2011).\\
S. D. Odintsov, and D. Saez-Gomez, Phys. Lett. B 725 (2013).\\
Z. Haghani, T. Harko, F. S. N. Lobo, H. R. Sepangi, and S. Shahidi, Phys. Rev. D 88, no. 4, 044023 (2013).\\
N. Tamanini, and T. S. Koivisto, Phys. Rev. D 88, no. 6, 064052 (2013).\\
F. G. Alvarenga, A. de la Cruz-Dombriz, M. J. S. Houndjo, M. E. Rodrigues, and D. Saez-Gomez, Phys. Rev. D 87 (2013).
\bibitem{23}
S. Nojiri, S. D. Odintsov, and M. Sasaki, Phys. Rev. D 71, 123509 (2005).\\
S. Nojiri, and S. D. Odintsov, Phys. Lett. B 631, 1 (2005).\\
G. Cognola, E. Elizalde, S. Nojiri, S. D. Odintsov, and S. Zerbini, Phys. Rev. D 73, 084007 (2006).\\
G. Calcagni, B. de Carlos, and A. De Felice, Nucl. Phys. B 752, 404 (2006).\\
T. Koivisto and D. F. Mota, Phys. Rev. D 75, 023518 (2007).\\
B. M. Leith, and I. P. Neupane, JCAP 0705, 019 (2007).\\
A. de la Cruz-Dombriz, and D. Saez-Gomez, Class. Quant. Grav. 29 (2012).
\bibitem{24}
E. Elizalde, R. Myrzakulov, V. V. Obukhov and D. Saez-Gomez, Class. Quant. Grav. 27, 095007 (2010).\\
R. Myrzakulov, D. Saez-Gomez and A. Tureanu, Gen. Rel. Grav. 43, 1671 (2011).
\bibitem{25}
P. Kanti, R. Gannouji, and N. Dadhich, Phys. Rev. D 92, 4, 041302 (2015).\\
P. Kanti, R. Gannouji, and N. Dadhich, Phys. Rev. D 92, 8, 083524 (2015).\\
S. Lahiri, JCAP 1609, 09, 025 (2016).\\
G. Hikmawan, J. Soda, A. Suroso, and F. P. Zen, Phys. Rev. D 93, 6, 068301 (2016).\\
M. Satoh, JCAP 1011, 024 (2010).\\
Z. K. Guo, and D. J. Schwarz, Phys. Rev. D 80, 063523 (2009).\\
S. Koh, B. H. Lee, W. Lee, and G. Tumurtushaa, Phys. Rev. D 90, no. 6, 063527 (2014).
\bibitem{26}
V. K. Oikonomou, Phys. Rev. D 92, 12, 124027 (2015).
\bibitem{27}
A. D. Linde, Phys. Lett. B108, 389 (1982).
\bibitem{28}
A. Albrecht and P.J. Steinhardt, Phys. Rev. Lett. 48, 1220 (1982).
\bibitem{29}
W. H. Press, Phys. Scr.21, 702 (1980).
\bibitem{30}
S. W. Hawking, Phys. Lett. B115, 295 (1982).
\bibitem{31}
A. A. Starobinsky, Phys. Lett. B117, 175 (1982).
\bibitem{32}
A. H. Guth, and S.Y. Pi, Phys. Rev. Lett. 49, 1110 (1982).
\bibitem{33}
J. M. Bardeen, P.J. Steinhardt, and M.S. Turner, Phys. Rev. D28, 679 (1983).
\bibitem{34}
V. F. Mukhanov, and G.V. Chibisov, JETP Lett. 33, 532 (1981).
\bibitem{35}
B. Whitt, Phys. Lett. B145, 176 (1984).
\bibitem{36}
D. Wands,  Class. Quant. Grav. 11, 269 (1994).
\bibitem{37}
A. K. Sanyal, Phys. Lett. B624, 81 (2005).
\bibitem{38}
A. K. Sanyal, Mod. Phys. Lett. A25, 2667 (2010).
\bibitem{39}
N. Sk, and A.K. Sanyal, J. Astrophys. Article ID 590171 (2013).
\bibitem{40}
K. Sarkar, N. Sk, S. Debnath, and A.K. Sanyal, Int. J. Theor. Phys. 52, 1194 (2013).
\bibitem{41}
B. Tajahmad, A. K. Sanyal, Eur. Phys. J. C, 77:217 (2017).
\bibitem{42}
H. Ooguri and C. Vafa, Nucl. Phys. B 766, 21-33 (2007).
\bibitem{43}
N. Arkani-Hamed, L. Motl, A. Nicolis and C. Vafa, JHEP 06, 060 (2007).
\bibitem{44}
M. Orellana, F. Garcia, F. Teppa Pannia and G. Romero, Gen. Rel. Grav 45, 771-783 (2013).
\bibitem{45}
K. Kadota, Ch. Sub Shin, T. Terada, and G. Tumurtushaa, JCAP01 008 (2020).\\
V. K. Oikonomou, arXiv:2012.01312 (2020).
\bibitem{46}
S. Capozziello, M. De Laurentis, S. D. Odintsov and A. Stabile, Phys. Rev. D 83, 064004 (2011).
\bibitem{a}
S. K. Garg, C. Krishnan, JHEP 11  075 (2019).
\bibitem{b}
S. Brahma, Md. Wali Hossain, JHEP 03  006 (2019).
\bibitem{c}
A. Ashoorioon, Physics Letters B, 790, 568-573 (2019).
\bibitem{d}
A. Achucarro, and G. A. Palma,  JCAP02 041 (2019).
\bibitem{e}
V. K. Oikonomou, Phys. Rev. D 103, 124028 (2021).
\bibitem{f}
C. Damian, O. Loaiza‐Brito, Fortsch.Phys. 67  1-2, 1800072 (2019).
\bibitem{g}
Chia-Min Lin JCAP06 015 (2020).
\bibitem{h}
Chien-I Chiang, J. M. Leedom, and H. Murayama, Phys. Rev. D 100, 043505 (2019).
\bibitem{i}
R. Adhikari, et al., The European Physical Journal C80, 899 (2020).
\bibitem{j}
R. Bravo, et al., JCAP02 004 (2020).
\bibitem{47}
O. Trivedi, arXiv:2101.00638 (2021).\\ S. Das, physics of the dark univers 27, 100432 (2020).
\bibitem{48}
A. Arapoglu, C. Deliduman and K. Y. Eksi, JCAP 1107, 020 (2011).
\bibitem{49}
A. Mohammadi, T. Golanbari, J. Enayati, arXiv:2012.01512, (2020).
\bibitem{50}
S. Capozziello, R. D'Agostino, O. Luongo, JCAP 1805, 008 (2018).
\bibitem{51}
C. Osses, N. Videla, and P. Grigoris, arXiv:2101.08882, (2021).
\bibitem{52}
S. Brahma, Phys.Rev.D 101  2, 023526 (2020).\\ R. Brandenberger,  arXiv:2102.09641 (2021).
\bibitem{53}
S. Capozziello, R. D'Agostino, O. Luongo, Gen. Rel. Grav. 51, 2 (2019).
\bibitem{54}
J. Sadeghi, E. Naghd Mezerji and S. Noori Gashti, Mod. Phys. Lett A. 36, 05, 2150027 (2021).
\bibitem{55}
R. Myrzakulov, L. Sebastian and S. Vagnozzi, Eur. Phys. J.C 75, 444 (2015).
\bibitem{56}
S. D. Odintsov and V. K. Oikonomou, Phys. Lett. B 805 (2020).
\bibitem{57}
S. D. Odintsov and V. K. Oikonomou, EPL126 no.2, 20002  (2019).
\bibitem{58}
S. D. Odintsov, V. K. Oikonomou and L. Sebastiani, Nucl. Phys. B, 923, 608 (2017).
\bibitem{59}
J. Sadeghi, S. Noori Gashti, and E.Naghd Mezerji. Phys. Dark Univ 30, 100626, (2020).\\
J. Sadeghi, Saeed Noori Gashti, Pramana 95(4) (2021).
\bibitem{60}
J. Sadeghi, S. Noori Gashti, E. Naghd Mezerjia, B. Pourhassan, arXiv:2011.05109 (2020).
\bibitem{61}
J. Sadeghi, E. Naghd Mezerji, S. Noori Gashti, arXiv:2011.14366 (2020).\\
S. Noori Gashti, Journal of Holography Applications in Physics 2 (1), 13-24 (2022).\\
J. Sadeghi, B. Pourhassan, Saeed Noori Gashti, and Sudhaker Upadhyay, Physica Scripta 96 (12), 125317 (2021).
\bibitem{62}
W. H. Kinney, Phys. Rev. Lett. 122, 8 081302 (2019).\\W. H. Kinney, arXiv:2103.16583 (2021).\\ Yu, Ten-Yeh and Wen, Wen-Yu, Phys. Lett. B781, 713-718  (2018).\\ V. K. Oikonomou, Phys. Rev. D 103, 124028(2021).
\bibitem{63}
J. Sadeghi, S. Noori Gashti, Eur. Phys. J. C 81, 301 (2021).
\bibitem{64}
M. v. Beest, J. Calderon-Infante, D. Mirfendereski, I. Valenzuela, arXiv:2102.01111 (2021).
\bibitem{65}
E. W. Kolb, A. J. Long, and E. McDonough, arXiv:2103.10437 (2021).
\bibitem{66}
A. Strominger and C. Vafa, Phys. Lett.B 379, 99-104 (1996).
\bibitem{67}
C. Vafa, arXiv:hep-th/0509212 (2005).
\bibitem{aa}
S. Das, Phys. Rev. D 99 6, 063514 (2019).\\
M. Shokri, J. Sadeghi, R. Herrera, S. N. Gashti, arXiv:2112.12309 (2021).
\bibitem{68}
A. De Simone, M.P. Hertzberg, F. Wilczek, Phys. Lett. B678, 1 (2009).
\bibitem{69}
S. C. Park, AIP Conf. Proc. 1078, 524 (2009).
\bibitem{70}
S. C. Park, S. Yamaguchi, JCAP 0808, 009 (2008).
\bibitem{71}
A. K. Sanyal, Phys. Lett. B624  81-92 (2005).
\bibitem{72}
A. K. Sanyal, Mod. Phys. Lett. A 25, 2667-2676 (2010).
\bibitem{73}
N. Sk, A. K. Sanyal, J. Astrophys. 2013  590171 (2013).
\bibitem{74}
S. Carloni, J.A. Leach, S. Capozziello, P.K.S. Dunsby, Class. Quantum Gravity 25, 035008 (2008).
\bibitem{75}
Y. Fujii, and K. Maeda in The Scalar-Tensor Theory of Gravity (Cambridge University Press, Cambridge, (2003).\\
N. D. Birrell, and P. C. W. Davies in Quantum Fields in Curved Space (Cambridge University Press, Cambridge, 1982).
\bibitem{76}
T. Rudelius, JCAP 08, 009 (2019).\\ H. Matsui and F. Takahashi, Phys. Rev. D 99, 023533 (2019).\\
K. Dimopoulos, Phys. Rev. D 98, 123516 (2018).
\bibitem{77}
S. Brahma, and S. Shandera, JHEP 11  016 (2019).
\bibitem{78}
T. Rudelius, JCAP 08  009 (2019).\\
\bibitem{79}
Z. Wang, R. Brandenberger, and L. Heisenberg, Eur. Phys. J. C 80  9, 864 (2020).
\bibitem{80}
J. Yuennan, and Ph. Channuie, Fortsch. Phys. 2200024 (2022).
\end{thebibliography}
\end{document}